\documentclass{article}

\usepackage{arxiv}

\usepackage[utf8]{inputenc} 
\usepackage[T1]{fontenc}    
\usepackage{hyperref}       
\usepackage{url}            
\usepackage{booktabs}       
\usepackage{amsfonts}       
\usepackage{nicefrac}       
\usepackage{microtype}      
\usepackage{lipsum}
\usepackage{graphicx}
\graphicspath{ {./images/} }

\usepackage{amsthm}
\usepackage{amssymb}
\usepackage{amsmath}
\usepackage[numbers,sort&compress]{natbib}
\usepackage{tabularx}
\usepackage{url} 

\theoremstyle{definition}
\newtheorem{assumption}{Assumption}
\newtheorem{remark}{Remark}
\newtheorem*{example}{Example}

\newenvironment{assumptionstar}[1][]{%
  \begin{assumption}[#1]%
}{%
  \end{assumption}%
}

\newcommand{\logit}{\text{logit}}

\newcommand{\vecWAX}{\begin{bmatrix}1&W&A&X\end{bmatrix}}
\newcommand{\hattheta}{\hat\theta}
\newcommand{\indep}{\perp\!\!\!\perp}

\title{Demystifying Proximal Causal Inference}
\author{
Grace V. Ringlein \\
  Department of Biostatistics\\
  Johns Hopkins Bloomberg School of Public Health\\
  Maryland, USA \\
  \texttt{gringle1@jh.edu} \\
   \And
 Trang Quynh Nguyen \\
  Department of Mental Health\\
  Johns Hopkins Bloomberg School of Public Health\\
  Maryland, USA  \\
  \texttt{trang.nguyen@jhu.edu} \\
  \And
 Peter P. Zandi \\
  Department of Psychiatry and Behavioral Sciences\\
  Johns Hopkins School of Medicine\\
  Maryland, USA \\
  \texttt{pzandi1@jhu.edu} \\
  \And
   Elizabeth A. Stuart \\
    Department of Biostatistics\\
  Johns Hopkins Bloomberg School of Public Health\\
  Maryland, USA \\
   \texttt{estuart@jhu.edu} \\
   \And
   Harsh Parikh \\
   Department of Biostatistics \\
   Yale University \\
   Connecticut, USA\\
   \texttt{harsh.parikh@yale.edu} \\
}

\begin{document}
\maketitle
\begin{abstract}
Proximal causal inference (PCI) has emerged as a promising framework for identifying and estimating causal effects in the presence of unobserved confounders. While many traditional causal inference methods rely on the assumption of no unobserved confounding, this assumption is likely often violated. PCI addresses this challenge by relying on an alternative set of assumptions regarding the relationships between treatment, outcome, and auxiliary variables that serve as proxies for unmeasured confounders. We review existing identification results, discuss the assumptions necessary for valid causal effect estimation via PCI, and compare different PCI estimation methods.  We offer practical guidance on operationalizing PCI, with a focus on selecting and evaluating proxy variables using domain knowledge, measurement error perspectives, and negative control analogies. Through conceptual examples, we demonstrate tensions in proxy selection and discuss the importance of clearly defining the unobserved confounding mechanism. By bridging formal results with applied considerations, this work aims to demystify PCI, encourage thoughtful use in practice, and identify open directions for methodological development and empirical research.
\end{abstract}


\textit{Note: This manuscript is a preprint and has not been peer reviewed.}

\section{Introduction}\label{sec_intro}
Across disciplines, researchers seek to answer causal ``what-if'' questions, such as comparing patients' response to different treatments or estimating the effect of a job training program on participants’ future earnings. While randomized experiments remain the gold standard to answer causal questions, they are often infeasible due to ethical, logistical, or financial constraints, and may not reflect real world implementation or broader populations. This leads researchers to analyze observational data, where a central challenge is confounding: factors that influence both treatment and outcomes, biasing naive comparisons. Traditional causal inference approaches typically assume all confounders are observed and measured without error---an assumption rarely met in practice. This motivates the question: \textit{can we identify causal effects even when some confounders are unmeasured, but some related proxy variables are observed?}

\begin{figure}
    \centering
    \includegraphics[width=0.5\linewidth]{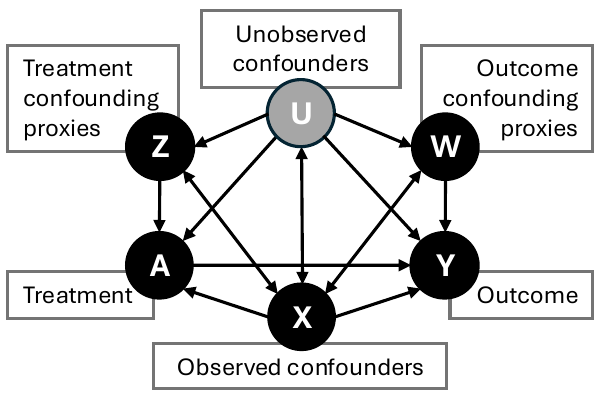}
    \caption{An example of a directed acyclic graph (DAG) illustrating the relationships between the two types of proxy variables and the unobserved confounders, treatment, and outcome. Arrows represent \textit{allowed} dependence, while a lack of arrow represents an independence assumption. For example, the lack of arrow between $Z$ and $Y$ represents the assumption that $Z\indep Y\mid U,A,X$ (A.\ref{condZ}). However, the arrow from $Z$ to $A$ reflects that $Z$ may or may not be a cause of $A$. Note that this is not the only DAG that can also satisfy PCI assumptions. See Supplemental Table A.1 in \citet{tchetgen_tchetgen_introduction_2024} for examples of other DAGs that do (and do not) satisfy PCI assumptions.} 
    \label{fig:pci_dag_labelled}
\end{figure}

The proximal causal inference (PCI) framework, introduced in \citet{miao_identifying_2018}, addresses unobserved confounding by leveraging two types of proxy variables: \textit{treatment confounding proxies} (related to outcomes only through confounders) and \textit{outcome confounding proxies} (related to treatments only through confounders) (see Figure~\ref{fig:pci_dag_labelled}). Building on measurement error correction \citep{kuroki_measurement_2014} and negative control methods \citep{lipsitch_negative_2010}, the key innovation of \citep{miao_identifying_2018} is \textit{nonparametric identification} of causal effects (meaning they can be written as a function of the observed data distribution without parametric assumptions) under less stringent assumptions than earlier work (e.g. \citep{kuroki_measurement_2014,carroll_measurement_2006}). Since its introduction, PCI has rapidly evolved to address longitudinal settings \citep{tchetgen_tchetgen_introduction_2024,ying_proximal_2021}, conditional average treatment effects \citep{sverdrup_proximal_2023}, selection bias \citep{li_doubly_2022}, continuous treatments \citep{wu_doubly_2023}, conditioning on post-treatment events \citep{park_proximal_2025}, synthetic controls \citep{liu_proximal_2023}, and text-based proxies \citep{chen_proximal_2024}. The proximal framework has also been extended to address unobserved mediators \citep{ghassami_causal_2023} and unobserved treatments \citep{zhou_causal_2024}, though in this work we focus explicitly on the original use case, unobserved confounders.

PCI is not the first or only framework in causal inference to address settings with possible unobserved confounding, but the goals, perspectives, and settings of other methods generally differ. \textit{Sensitivity analyses} for unobserved confounding aim to quantify how strong an unobserved confounder would have to be to change significance of results \citep{cornfield_smoking_1959,rosenbaum_assessing_1983,liu_introduction_2013}, and traditional negative control methods have typically aimed to \textit{detect} bias due to unobserved confounding, rather than correct it \citep{lipsitch_negative_2010,shi_selective_2020,yang_advances_2024}. The instrumental variables (IV) framework also addresses unobserved confounding but utilizes a different type of auxiliary variable, called an \textit{instrument}: an exogenous variable relevant to the exposure of interest, that must be independent of unobserved confounders (conditional on observed covariates) \citep{angrist_identification_1996,baiocchi_instrumental_2014}. In PCI, the proxies must \textit{not} be independent of the unobserved confounder (conditional on observed covariates and treatment assignment).

Despite its theoretical elegance, applying PCI methods may be challenging. PCI generally relies on technical concepts such as bridge functions that solve integral equations and completeness conditions which may not be covered in standard statistics curricula. Special cases---such as when the proxies and outcome are all normally distributed \citep{tchetgen_tchetgen_introduction_2024, zivich_introducing_2023} or when data are fully categorical \citep{shi_multiply_2020}---may simplify identification and estimation. However, in real-world settings with more complex data structures, understanding the implications of PCI assumptions and which estimation methods are appropriate can be challenging. While applied examples do appear in the PCI literature, they rarely engage deeply with the plausibility of the full set of underlying assumptions. 

This paper aims to \textit{demystify} PCI by providing: (1) a structured overview of PCI, covering assumptions, identification strategies, and estimation methods; (2) practical guidance on selecting proxies (emphasizing domain knowledge, conceptualization of confounding mechanisms, and connections to measurement error and negative controls); and (3) an illustration of tensions between PCI assumptions using a conceptual example. Our goal is to provide a comprehensive resource that facilitates thoughtful adoption and appropriate use of PCI in applied research, identify areas for future methodological development, and ultimately, contribute to more robust causal inferences in the presence of unmeasured confounders.

\section{Background: Causal Inference Under Unconfoundedness}\label{sec_prelim}

Throughout this work, $A$ denotes exposure or treatment (typically binary), $Y$ is the observed outcome, and $Y(a)$ denotes the potential outcome under treatment level $A=a$ \citep{rubin_estimating_1974}. The average treatment effect (ATE), defined as $\mathbb{E}[Y(1) - Y(0)]$ and denoted $\tau$ is a common target of causal inference and is our focus, though much of the discussion is relevant to other causal estimands. 

The majority of standard non-experimental causal inference methods rely on three key assumptions utilizing a set of observed covariates $X$ (typically measured prior to treatment assignment):

\begin{assumption}[Consistency]\label{cons}
$Y = Y(1)A + Y(0)(1-A).$
\end{assumption}

\begin{assumption}[Positivity]\label{pos}
$0 < \mathbb{P}(A=a \mid X), \text{ for } a \in \{0,1\}.$
\end{assumption}

\begin{assumption}[Unconfoundedness]\label{exc}
$Y(a) \indep A \mid X \text{ for } a \in \{0,1\}.$
\end{assumption}

\textit{Consistency} (A.\ref{cons}) requires that treatment assignment determines which potential outcome is observed, without altering the potential outcomes\footnote{This is essentially equivalent to \textit{stable unit-treatment value assumption} which is also commonly invoked in causal interference literature \citep{rubin_randomization_1980}.}, while \textit{positivity} (A.\ref{pos}) means that within every level of $X$ there is a non-zero probability for units to end up in either condition. \textit{Unconfoundedness} (A.\ref{exc}), also known as \textit{exchangeability},
 essentially claims that, after conditioning on $X$, treatment assignment is ``as good as random.'' Unconfoundedness is fundamentally untestable and often viewed skeptically in observational studies, as researchers can never be certain they have measured all relevant confounders. PCI's key innovation to provides identification results without requiring unconfoundedness, though it introduces other assumptions (Section~\ref{sec_pci}), many of which are also untestable.

Under A.\ref{cons}, A.\ref{pos}, and A.\ref{exc}, the mean of $Y(a)$ is identified, which can be expressed in different ways, such as:
\begin{align*}
    &\mathbb{E}[Y(a)]\overset{\text{ID}}{=}\mathbb{E}\{\mathbb{E}[Y\mid X,a]\} =\mathbb{E}\left[\frac{\mathbb{I}(A=a)}{\mathbb{P}(A\mid X)}Y\right]=\mathbb{E}\left[\frac{\mathbb{I}(A=a)}{\mathbb{P}(A\mid X)}\{Y-\mathbb{E}[Y\mid X,a]\} +\mathbb{E}[Y\mid X,a]\right].
\end{align*}
 
Examples of ATE estimators justified by these expressions are: 
\begin{enumerate}
    \item The outcome-regression estimator $\hat\tau_{or}=\mathbb{P}_n[\hat\mu_1(X)-\hat\mu_0(X)]$ (where $\hat\mu_a(X)$ indicates prediction based on a regression model that estimates $\mathbb{E}[Y\mid X,A=a]$ and $\mathbb{P}_n$ denotes averaging over the sample);
    \item The inverse probability weighting (IPW) estimator $\hat\tau_{ipw}=\mathbb{P}_n\left[\frac{AY}{\hat{e}(X)}-\frac{(1-A)Y}{1-\hat{e}(X)}\right]$ (where $\hat{e}(X)$ is the so-called propensity score model, which estimates $\mathbb{P}(A=1\mid X)$); and
    \item The augmented IPW estimator (AIPW) $\hat\tau_{aipw}=\mathbb{P}_n\left[\frac{A[Y-\hat\mu_1(X)]}{\hat{e}(X)}+\hat\mu_1(X)-\frac{(1-A)[Y-\hat\mu_0(X)]}{1-\hat{e}(X)}-\hat\mu_0(X)\right]$ (which is doubly robust, i.e., consistent if either the outcome regression or the propensity score model is correctly specified, but not necessarily both) \citep{bang_doubly_2005}. 
\end{enumerate}  
These standard estimators have direct parallels to PCI estimators that will be discussed in Section \ref{sec_estimation}. 

These and other estimators derived under the assumptions A.\ref{cons}, A.\ref{pos}, and A.\ref{exc} may be biased when one or more of these assumptions are violated---our focus in this work is violation of A.\ref{exc}. 

\section{Proximal Causal Inference Framework}\label{sec_pci}
\subsection{PCI Assumptions}\label{sec_base_assumptions}

PCI addresses scenarios where unconfoundedness (A.\ref{exc}) is suspect. \textit{Latent unconfoundedness} (A.\ref{latent}) is assumed instead of A.\ref{exc}, along with consistency (A.\ref{cons}) and a modified positivity condition (A.\ref{latpos}), which is similar to A.\ref{pos} but conditions on both $U$ and $X$.

\begin{assumption}[Latent Unconfoundness]\label{latent}
$Y(a) \indep A \mid U,X \text{ for } a \in \{0,1\}.$
\end{assumption}

\begin{assumption}[Latent Positivity]\label{latpos}
$0 < \mathbb{P}(A=a \mid X, U) \text{ for } a \in \{0,1\}.$
\end{assumption} 

\begin{figure}
    \centering
    \includegraphics[width=0.33\linewidth]{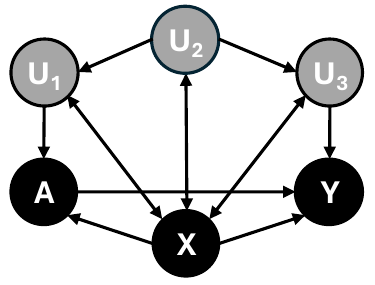}
    \caption{A example causal diagram where there are multiple ways to define a set of unobserved confounders $U$ that satisfy $Y(a) \indep A \mid U,X$ (latent unconfoundedness, A.\ref{latent}): as any of the three unobserved variables individually ($U=U_1,U=U_2,$ or $U=U_3$), any pair of them (e.g. $U=\{U_1,U_2\}$), or all three ($U=\{U_1,U_2,U_3\}$).}
    \label{fig:u1u2u3}
\end{figure}

Assumption A.\ref{latent} (also known as latent exchangeability \citep{tchetgen_tchetgen_introduction_2024}) means there exists a set of unobserved variables $U$ such that if we had observed them, we would have unconfoundedness in the usual sense. Note that in some cases, there may be multiple ways to define a set of variables $U$ that could satisfy A.\ref{latent}---see Figure \ref{fig:u1u2u3} for an example. 

Trivially, if $U = \emptyset$ satisfies A.\ref{latent}, then traditional unconfoundedness (A.\ref{exc}) holds. In this work, we are generally concerned with scenarios where it is reasonable to believe $U\not=\emptyset$ would not satisfy A.\ref{latent}. Otherwise, use of PCI would not be motivated. 

\begin{remark}\label{remark_udim}
    While the assumption that \textit{there exists} a set of variables $U$ that would satisfy A.\ref{latent} is generally reasonable by itself---as $U$ could consist of an extremely large number of variables---the challenge is that subsequent assumptions that depend on $U$ must also hold. It may be more difficult to make some of these assumptions (i.e., the completeness conditions that will be discussed in Section \ref{sec_idh}) when $U$ is extremely complex. Having a specific conceptualization of $U$ (such as ``health seeking behavior" \citep{shi_multiply_2020}) is integral to being able to reason about the full set of PCI assumptions. 
\end{remark}

Under these three assumptions alone (A.\ref{cons}, A.\ref{latent}, A.\ref{latpos}), the potential outcome means and ATE are not identified due to $U$ being unobserved. PCI requires additional assumptions that use two different types of proxies for $U$:  \textit{treatment confounding proxies} (denoted $Z$) and \textit{outcome confounding proxies} (denoted $W$), defined by A.\ref{condZ} and A.\ref{condW} below:

\begin{assumption}[Treatment Confounding Proxy ($Z$) Conditional Independence]\label{condZ}
$Z \indep Y \mid U,X,A.$
\end{assumption}

\begin{assumption}[Outcome Confounding Proxy ($W$) Conditional Independence]\label{condW}
$W \indep (A,Z) \mid U,X.$
\end{assumption} 

\begin{table*}[t]
\caption{Base Assumptions and Necessary Conditions for Proximal Causal Inference}
\label{tab:core}
\setlength{\tabcolsep}{6pt}
\renewcommand{\arraystretch}{1.5}

\begin{tabularx}{16cm}{
  @{}
  >{\raggedright\arraybackslash}p{4cm}
  >{\raggedright\arraybackslash}p{4cm}
  >{\raggedright\arraybackslash}p{7cm}
  @{}
}
\toprule
\textbf{Assumption or Condition Name} & \textbf{Formal statement} & \textbf{Plain-language restatement} \\
\midrule

Consistency (A.\ref{cons}) & 
$Y = A Y(1) + (1-A)Y(0)$ & 
The observed outcome equals the potential outcome under the treatment actually received. \\

Latent unconfoundedness (A.\ref{latent}) &
$Y(a) \perp A \mid U, X$ & 
Given the unobserved confounder $U$ and observed covariates $X$, treatment assignment would be as good as random. \\

Latent positivity (A.\ref{latpos}) &
$0 < P(A=a \mid U, X)$ & 
For every level of $U$ and $X$, each treatment has a positive probability of being received. \\

Conditional independence assumption for treatment confounding proxies (A.\ref{condZ}) & 
$Z \indep Y \mid U, A, X$ & 
Once we condition on the unobserved confounder $U$, observed covariates $X$, and treatment assignment $A$, treatment confounding proxies $Z$ carry no extra information about the outcome $Y$. \\

Conditional independence assumption for outcome confounding proxies (A.\ref{condW}) & 
$W \indep (A, Z) \mid U, X$ & 
Once we condition on the unobserved confounder $U$ and observed covariates $X$, outcome confounding proxies $W$ carry no extra information about treatment assignment $A$ or $Z$. \\

$U$-relevance of treatment confounding proxies &
 $Z \not\indep U \mid A,X$ & 
For every level of $A$ and $X$, $Z$ carries information about $U$.
 \\

$U$-relevance of outcome confounding proxies&
 $W \not\indep U \mid A,X$ & 
For every level of $A$ and $X$, $W$ carries information about $U$.\\

\bottomrule
\end{tabularx}
\end{table*}

Intuitively, A.\ref{condZ} means that the treatment confounding proxies ($Z$) must not be linked to the outcome ($Y$) beyond shared association with $U, A, X$. Similarly, A.\ref{condW} requires that the outcome confounding proxies ($W$) should not be linked to the treatment assignment ($A$) or treatment confounding proxies ($Z$) other than through their shared association with $U$ and $X$. Figure \ref{fig:pci_dag_labelled} is an  example of a DAG that satisfies these assumptions--- see the supplement of \citet{tchetgen_tchetgen_introduction_2024}, Table A.1 for additional examples. 

Additional assumptions about how $Z$ and $W$ relate to $U$ will be used for identification. While the formal assumptions (e.g., \textit{completeness conditions}) technically differ between the two main paths to identification \citep{miao_identifying_2018,tchetgen_tchetgen_introduction_2024,cui_semiparametric_2024} (discussed in Sections \ref{sec_idh} and \ref{sec_idq}), necessary conditions for identification in either case are that $Z$ and $W$ are $U$-relevant, meaning that both sets of proxies must be associated with $U$ (i.e. not independent of $U$) conditional on $A,X$.

As an example of a treatment confounding proxy and outcome confounding proxy pair used in a PCI analysis, consider comparing adverse event risk ($Y$) after a novel vaccination ($A$) \citep{shi_multiply_2020}. \citet{shi_multiply_2020} used ringworm-related visits as a treatment confounding proxy ($Z$) and post-vaccination injury visits as an outcome confounding proxy ($W$) for unmeasured health-seeking behavior ($U$). Injury visits should be unrelated to vaccine choice except through health-seeking, and vaccine choice is unlikely to be related to ringworm diagnosis other than through health-seeking \citep{shi_multiply_2020}. 

In the following remarks, we provide some clarifications about these proxies and their role in PCI. Section~\ref{sec_practical} elaborates further on several of these points, including the conceptual similarities between between proxies, error-prone measurement, and negative controls and strategies for proxy selection. 

\begin{remark}\label{remark_need_both}
To point identify and estimate the ATE in most PCI methods, it is necessary to have at least one observed variable that can be used as a treatment confounding proxy, and at least one other observed variable that can be used as an outcome confounding proxy. That said, when only one type of proxy appears to be available in a particular application, it may still be possible to \textit{partially identify} (obtain bounds on) the average treatment effect \citep{ghassami_partial_2023} or to identify a different estimand \citep{park_single_2024}---however, that is not the focus of this work. 
\end{remark}

\begin{remark}\label{remark_disconnected} 
Treatment confounding proxies ($Z$) are allowed to be associated with treatment assignment $A$ conditional on $U,X$, but this is not required. Similarly, outcome confounding proxies ($W$) need not be associated with $Y$ conditional on $U,X$. Thus, a special case is a \textit{disconnected} proxy: a variable that is believed to be independent of \textit{both} $A$ and $Y$ conditional on $U$, $X$. Such a variable is good candidate for use as either a treatment confounding proxy or an outcome confounding proxy. When multiple disconnected proxies are available, the consistency of the PCI estimate should not depend on which disconnected proxies are classified as a treatment confounding proxy or an outcome confounding proxy---as long as the additional assumptions required are also valid \citep{shi_multiply_2020, tchetgen_tchetgen_introduction_2024}. Measurements in a traditional latent variable model with classical measurement error are examples of disconnected proxies. Further discussion of types of proxies in the language of measurement error can be found in Appendix~\ref{appendix_measurement}.

\end{remark}

\begin{remark}\label{remark_invalid_bias}
Incorrectly including a variable as a treatment confounding proxy or outcome confounding proxy that does not satisfy their respective assumptions (A.\ref{condZ}, A.\ref{condW}) may result in bias \citep{yu_fortified_2025,huang_relative_2025,rakshit_adaptive_2025}. Determining sets of proxies that reasonably satisfy A.\ref{condZ} and A.\ref{condW} (and subsequent additional assumptions) in a particular applied scenario relies primarily on domain knowledge and data availability, as $U$ is unobserved. 
\end{remark}

\begin{remark}\label{remark_names}
    Multiple names are used for the two types of proxies in PCI. Some works refer to them primarily as negative control exposures and outcomes \citep{shi_multiply_2020,xie_automating_2024,miao_confounding_2024}---we discuss this in Section~\ref{sec_practical}. \citet{tchetgen_tchetgen_introduction_2024} often refer to treatment confounding proxies and outcome confounding proxies as ``type b" and ``type c" variables (where observed confounders $X$ are ``type a"). Still others call them outcome and treatment \textit{inducing} proxies (rather than \textit{confounding} proxies) \citep{rakshit_adaptive_2025,cobzaru_bias_2022}, though we note that this terminology only makes sense for the case where $Z$ and $W$ are causes of $A$ and $Y$, respectively. \citet{mastouri_proximal_2023} often simply refers to them as $Z$ and $W$. The names that are currently most prominent in PCI, ``treatment confounding proxies" and ``outcome confounding proxies," reflect the \textit{allowed} dependence between proxies and outcome or treatment: there is \textit{no} restriction on whether the \textit{treatment} confounding proxies ($Z$) are dependent or independent of the \textit{treatment} ($A$, conditional on $U,X$), and \textit{outcome} confounding proxies ($W$) are \textit{not} restricted in their relationship to the \textit{outcome} ($Y$) conditional on $U,X$. We note that it could be more intuitive to refer to these proxies by referencing the relationships that are \textit{restricted} by assumptions A.\ref{condZ} and A.\ref{condW}, rather than those that are \textit{not} restricted: $Z$ could be referred to as ``outcome independent" or ``outcome disconnected" proxies, and the $W$ as ``treatment independent" or ``treatment disconnected" proxies. However, we do not attempt to rename these variables in this work, to aid the reader in connecting this work to the existing literature. 
\end{remark} 

\begin{remark}\label{remark_iv}
     Assumptions A.\ref{condZ} and A.\ref{condW} have similarity to the \textit{exclusion restriction} assumption used in the well-known instrumental variables (IV) framework \citep{angrist_identification_1996,baiocchi_instrumental_2014}. IV approaches assume the following exclusion restriction: an observed variable (termed ``instrument'') is associated with treatment but is independent of the outcome given the treatment and covariates; the typical case is where the instrument affects treatment assignment but does not affect the outcome (other than through the treatment) and is not connected to the outcome through unobserved causes. This connection has previously been recognized in the PCI literature, with some work \citep{yu_fortified_2025,rakshit_adaptive_2025} even referring to assumptions A.\ref{condZ} and A.\ref{condW} as ``exclusion restrictions". The key distinction is that IV validity for an instrumental variable (denoted $V$) requires $V \perp U \mid X$, whereas treatment confounding proxies in PCI must not be independent of $U$ conditional on $A,X$. However, when $V$ is a cause of $A$, conditioning on $A$ can induce dependence between $V$ and $U$ through collider stratification, potentially allowing $V$ to be complete for $U$ given $A,X$ \citep{shi_selective_2020, tchetgen_tchetgen_introduction_2024}. The practical implications of using a valid instrumental variable as a treatment confounding proxy in PCI have not yet been well explored, to our knowledge.

\end{remark}

Assumptions A.\ref{cons}, A.\ref{latpos}, A.\ref{latent}, A.\ref{condZ}, A.\ref{condW} are consistently invoked across PCI methods--- thus, we will refer to them as \textit{base} PCI assumptions. 
We also consider $U$-relevance (that $Z$ and $W$ must both be associated with $U$ conditional on $A,X$) as part of these base assumptions---though technically, these are necessary conditions of formal assumptions to come, rather than unique identification assumptions themselves. 

\begin{figure*}
    \centering
\includegraphics[width=1\linewidth]{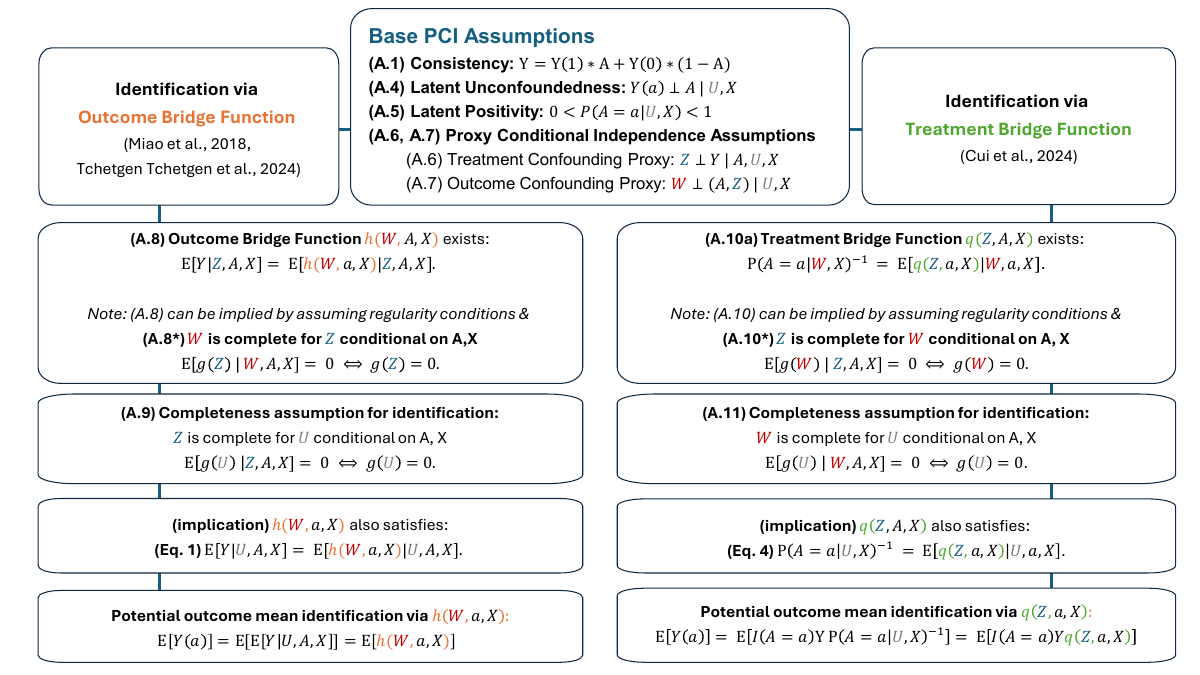}
    \caption{This diagram lists the core assumptions of PCI and summarizes the assumptions of two complementary identification approaches: (i) via the Outcome Bridge Function \citep{miao_identifying_2018,tchetgen_tchetgen_introduction_2024}, and (ii) via the Treatment Bridge Function \citep{cui_semiparametric_2024}. For each approach, the required bridge function assumption, completeness conditions, and formulas for average treatment effect (ATE) identification are displayed. Note that in both approaches, one completeness condition is required for identification, while another completeness condition can be used with regularity conditions \citep{miao_identifying_2018,cui_semiparametric_2024} to justify that a bridge function exists. Variations on these paths are presented in \citet{miao_confounding_2024} (see Remark~\ref{remark:miao2024}) and \citet{cui_semiparametric_2024}. }
    \label{fig:identification_paths}
\end{figure*}

Table~\ref{tab:core} summarizes these assumptions and Appendix~\ref{ex:combo} discusses how these assumptions map onto the coefficients of an example parametric model. Additional assumptions are required for identification and will be introduced in the next section, in the context of the two main paths to identification of the ATE in PCI, from \citet{miao_identifying_2018} and \citet{cui_semiparametric_2024} (summarized in Figure~\ref{fig:identification_paths}).

\subsection{Nonparametric Identification via an Outcome Bridge Function}\label{sec_idh}

This section presents the first of two identification strategies, based on an \textit{outcome bridge function} that links the outcome $Y$ to the outcome confounding proxy $W$ in a way that implicitly adjusts for 
$U$. A second identification 
strategy introduced in \citet{cui_semiparametric_2024}, based on a \textit{treatment bridge function}, is presented in Section~\ref{sec_idq}. Within the outcome bridge function identification ``path", there are variations in approach; we focus on the original strategy due to \citet{miao_identifying_2018} and \citet{tchetgen_tchetgen_introduction_2024}; but draw on the alternate approach of \citet{miao_confounding_2024} largely for intuition---similarities and differences between the identification approaches are discussed in Remark~\ref{remark:miao2024} at the end of this section. 

\subsubsection{Intuition for the Outcome Bridge Function}

At the core of this identification path is an \textit{outcome bridge function} $h(W,A,X)$ that satisfies the following relationship, for $a=0,1$ \citep{miao_identifying_2018,tchetgen_tchetgen_introduction_2024,miao_confounding_2024}.
\begin{align}\label{PCIhu}
    \mathbb{E}[Y\mid U,A=a,X]= \mathbb{E}[h(W,a,X)\mid U,A=a,X]
\end{align}

 Eq. \ref{PCIhu} is also commonly expressed as the equivalent integral (or summation) over the support of $W$ (given $u,a,x$) and thus is often referred to as an \textit{integral equation} \citep{miao_identifying_2018}: $$\mathbb{E}[Y\mid U,A=a,X]=\int_wh(w,a,X)f(W=w\mid U,a,X).$$
 
We defer the discussion of the existence of such an $h(W,A,X)$ function briefly, and begin here by discussing intuition behind the idea of such a function, following \citet{miao_confounding_2024}. Eq. \ref{PCIhu} means that for $A=a$ and any $X=x$, the effect of $U$ on $Y$ is equal to the effect of $U$ on some transformation $h(W,a,x)$ of $W$, hence the label \textit{outcome bridge function}. This requires $W$ to be $U$-relevant, i.e., associated with $U$ conditional on $A,X$ (otherwise the RHS does not vary with $U$ but the LHS does).%
\footnote{Assume for contradiction $W\indep U\mid A, X$. Then $\mathbb{E}[Y\mid U,A,X]=\mathbb{E}[h(W,A,X) \mid U,A,X]=\mathbb{E}[h(W,A,X) \mid A,X]$. But if $U\not=\emptyset$, $\mathbb{E}[Y\mid U,A,X]$ must vary with $U$---a contradiction.} Note that if $U$ were observed, we could utilize $\mathbb{E}[Y\mid U,X,A]$ directly for identification, $\mathbb{E}[Y(a)]=\mathbb{E}[\mathbb{E}[Y\mid U,X,A=a]]$ (under the consistency, positivity and unconfoundedness assumptions). With $U$ not observed, as long as function  $h(W,a,X)$ exists, it can stand in for $\mathbb{E}[Y\mid U,X,A=a]$. To see why, note that $h(W,a,X)$ is a function of $W$ rather than $U$, so for any fixed $u,a,x$ its value will vary with randomness in $W$. However, Eq.~\ref{PCIhu} would guarantee that this variation averages out: the conditional mean of $h(W,a,X)$ given $U,X,A=a$ is exactly $\mathbb{E}[Y\mid U,A=a,X]$. In this sense, $h(W,a,X)$ serves as a ``noisy version" of the unobservable quantity $\mathbb{E}[Y\mid U,X,A=a]$---one that has variability at the individual level (through its dependence on $W$) but recovers the correct conditional mean upon averaging.

The requirement that $W\indep A\mid U,X$ (part of A.\ref{condW}) simplifies the RHS of Eq. \ref{PCIhu}: $\mathbb{E}[h(W,a,X)\mid U,X,A=a]=\mathbb{E}[h(W,a,X)\mid U,X]$. This obtains a simple result for the potential outcome mean:
\begin{align*}
    \mathbb{E}[Y(a)]=\mathbb{E}\{\mathbb{E}[h(W,a,X)\mid U,X]\}=\mathbb{E}[h(W,a,X)].
\end{align*}

However, $h(W,A,X)$ is not generally known and must be learned from observed data---else this relationship provides little utility. We need an additional source of information about $U$, which is provided by the treatment confounding proxies, $Z$. Essentially, an observable counterpart of Eq. \ref{PCIhu} can be used, which incorporates $Z$ instead of $U$ \citep{miao_identifying_2018,tchetgen_tchetgen_introduction_2024,miao_confounding_2024}:
\begin{align}\label{PCIh:eq}
    \mathbb{E}[Y\mid Z,A,X]=\mathbb{E}[h(W,A,X)\mid Z,A,X].
\end{align}

The connection between Eq. \ref{PCIhu} and Eq. \ref{PCIh:eq} formalized in the identifying assumptions that we now introduce. 

\subsubsection{Identification results of \citet{miao_identifying_2018} and \citet{tchetgen_tchetgen_introduction_2024}}

The original identification results of \citet{miao_identifying_2018} and \citet{tchetgen_tchetgen_introduction_2024} begin by assuming an \textit{outcome bridge function} exists that solves Eq. \ref{PCIh:eq} above, restated as assumption A.\ref{PCIh}, and then use a completeness condition to show that any solution to this observable equation must also satisfy the $U$-level relationship in Eq.~\ref{PCIhu}.

\begin{assumption}
[Outcome Bridge Function]\label{PCIh}
There exists a function $h(W,A,X)$ that solves 
\begin{align}
\mathbb{E}[Y\mid Z,a,X]&= \mathbb{E}[h(W,a,X)\mid Z,a,X].\tag{\ref{PCIh:eq}}
\end{align}
\end{assumption}

\begin{assumption}[$Z$ is complete for $U$ conditional on $A,X$]\label{compUZ} For any $a,x$ and square-integrable function $g$, we have:\footnote{To say a function $g(U)$ of some random variable $U$ (with probability distribution $P_u$) is \textit{square integrable} means that $\int_\mathcal{U}|g(u)|^2dP_u<\infty$. Equivalently, $g(U)$ has finite second moment. See \citet{andrews_examples_2017} and references within for background on this specific type of completeness, known formally as $L^2$-completeness.}
$$\mathbb{E}[g(U) \mid Z,a,x] = 0\  a.s. \iff g(U) = 0.\  a.s. $$
\end{assumption}

A.\ref{compUZ} requires that for any particular $A=a,X=x$, the set of treatment confounding proxies, $Z$, has sufficient variability relative to the variability of $U$ (interpretation of completeness conditions is discussed further in Section~\ref{sec_completeness_intro}). Specifically, under the conditional independence assumptions (A.\ref{condZ}, A.\ref{condW}) and the fact that $h$ averages out ``correctly" for every value of $Z$ (A.\ref{PCIh}), completeness condition A.\ref{compUZ} is sufficient to conclude that $h$ also averages ``correctly" for every value of $U$---that is, Eq.~\ref{PCIhu} holds. This is the critical step: it promotes an observable integral equation into the $U$-level property (Eq.~\ref{PCIhu}). The counterfactual mean is then identified \citep{miao_identifying_2018,tchetgen_tchetgen_introduction_2024}:\footnote{This same result as \citet{miao_confounding_2024}.} 
\begin{align}\label{PCI_h_id}
\mathbb{E}[Y(a)] &\overset{\text{ID}}{=} 
\mathbb{E}[h(W,a,X)]
\end{align}
and the ATE can be expressed as:
\begin{align}
    \tau_{h}
    &=\mathbb{E}[h(W,1,X)  - h(W,0,X)]
\end{align}

This can be called the \textit{proximal g-formula} given its similarity to the standard g-formula identification of the ATE ($\tau=\mathbb{E}[\mu_1(X)-\mu_0(X)]$, where $\mu_a(X)=\mathbb{E}[Y\mid X,A=a]$) \citep{tchetgen_tchetgen_introduction_2024}. Thus, estimation approaches based on this result (i.e., $\hat\tau_h=P_n[\hat h(W,1,X)-\hat h(W,0,X)]$) may be referred to as \textit{proximal g-computation} \citep{tchetgen_tchetgen_introduction_2024}.

Instead of assuming A.\ref{PCIh} directly, an alternative is to assume an additional completeness condition: A.\ref{compZW}, below.\footnote{The definition of this and subsequent completeness condition with parallel forms to A.\ref{compUZ} will be abbreviated to reduce redundancy.}
Under regularity conditions provided in the Appendix of \citet{miao_identifying_2018} (as well as in \citet{cui_semiparametric_2024}), A.\ref{compZW} implies not only the existence of the outcome bridge function (A.\ref{PCIh}) but also its uniqueness. That said, uniqueness of the bridge function is not required for identification of the $\mathbb{E}[Y(a)]$ \citep{zhang_proximal_2023}: all solutions to Eq. \ref{PCIh:eq} yield the same potential outcome mean when integrated over the distribution of $W,X$. 

\addtocounter{assumption}{-2}%
\begin{assumptionstar}[$W$ is complete for $Z$ conditional on $A,X$]\label{compZW}
     $\mathbb{E}[g(Z) \mid W, A,X)] = 0 \iff g(Z) = 0$
\end{assumptionstar}
\addtocounter{assumption}{1}%

A more detailed outline of the identification results of \citet{miao_identifying_2018,tchetgen_tchetgen_introduction_2024} is provided in Appendix~\ref{appendix_h_ident}. 

\begin{remark}\label{remark:miao2024}
A variation on this identification approach is introduced in \citet{miao_confounding_2024}. It begins by assuming an outcome bridge function exists that satisfies Eq.~\ref{PCIhu}: a relationship involving the unobserved $U$. This allows derivation of Eq. \ref{PCIh:eq}: the observable counterpart of Eq. \ref{PCIhu}. This can be seen by taking the expectation of both sides of Eq.\ref{PCIhu} over $f(U\mid Z,A,X)$, invoking conditional independence assumptions A.\ref{condZ} and A.\ref{condW}, and using the definition of iterated expectation. Under a completeness condition (that $Z$ is complete for $W$ given $A,X$, A.\ref{compWZ}, which we introduced later in Section~\ref{sec_idq} in a different context) this bridge function is unique and the potential outcome mean is identified with same expression as in \citet{miao_identifying_2018} (Eq. \ref{PCI_h_id}). \citet{miao_confounding_2024} also present an alternative to this completeness condition for scenarios where a parametric form of the outcome bridge function satisfying Eq. \ref{PCIhu} is assumed. Thus, while both the identification results of \citet{miao_identifying_2018} and the identification approach in \citet{miao_confounding_2024} hinge on bridge function that satisfies both Eq. \ref{PCIhu} and Eq. \ref{PCIh:eq},  \citet{miao_confounding_2024} goes from Eq. \ref{PCIhu} to Eq. \ref{PCIh:eq}, the opposite direction of \citet{miao_identifying_2018}, which goes from Eq. \ref{PCIh:eq} to Eq. \ref{PCIhu} (using different completeness assumptions). The approach of \citet{miao_identifying_2018} requires the bridge function satisfying \ref{PCIh:eq} to be unique (using A.\ref{compWZ}) for identification, while the approach of \citep{miao_identifying_2018,tchetgen_tchetgen_introduction_2024} does not---though non-uniqueness must then be addressed in estimation and inference \citep{zhang_proximal_2023}. While both approaches utilize the same core machinery original approach \citep{miao_identifying_2018,tchetgen_tchetgen_introduction_2024} has the advantage of starting from an equation that can be estimated from observed data, while the alternate approach \citep{miao_confounding_2024} provides more direct intuition about the outcome bridge function's role in identification. 
\end{remark}

\begin{remark}\label{remark_fredholm}
    The form of Eq. \ref{PCIh:eq} in A.\ref{PCIh} (and of Eq. \ref{PCIhu}), when written as an integral, is known as a \textit{Fredholm integral equation of the first kind}. A challenge to the accessibility of PCI is that such equations, while fundamental in graduate-level (applied) mathematics, are not typically covered in standard statistics curricula. A key point is that these equations are often \textit{ill-posed}: a solution may not exist, may exist but not be unique, or may fail to change smoothly with changes in the data (i.e., it is unstable) \citep{kress_linear_2014}. This leads directly to challenges in estimation \citep{zhang_proximal_2023}.
\end{remark}

\begin{remark}\label{remark_known_forms}
If the form of $\mathbb{E}[Y\mid Z, A,X]$ and $f(w\mid Z,A,X)$ were known, then in theory, factorization (e.g. via singular value decomposition) could be used to identify the bridge function \citep{kress_linear_2014}. Practically, such factorization might lead to a solution without a closed form for the bridge function which may be hard to operationalize. Thus, it is common to assume a functional form for the outcome bridge function (e.g. $h(W,A,X)=\eta_wW + \eta_aA + \eta_xX$) and use that for estimation rather than attempting to solve for $h(W,A,X)$ from some assumed $\mathbb{E}[Y\mid A,X,Z]$ and $f(w\mid Z,A,X)$ \citep{tchetgen_tchetgen_introduction_2024,cui_semiparametric_2024}. However, it is possible that that functional form might be misspecified and can lead to bias in estimation \citep{cui_semiparametric_2024}. This applies equivalently to the treatment bridge function used in the second identification result discussed in Section~\ref{sec_idq}. 
\end{remark}

\subsection{Nonparametric Identification via a Treatment Bridge Function}\label{sec_idq}

\subsubsection{Intuition for the Treatment Bridge Function}

The identification result of \citet{cui_semiparametric_2024} has a parallel structure to the outcome bridge function results. Where the outcome bridge function $h(W,A,X)$ 
serves as a ``noisy" version of $\mathbb{E}[Y\mid U,A,X]$, the \textit{treatment bridge function} $q(Z,A,X)$ serves as a ``noisy" version of the inverse probability of treatment, $\mathbb{P}(A=a\mid U,X)^{-1}$:
\begin{align}\label{PCIqu}
    \mathbb{P}(A=a\mid U,X)^{-1} &=\mathbb{E}[q(Z,a,X)\mid U,A=a,X].
\end{align}

Following the structure of the intuition provided by \citet{miao_confounding_2024} for the outcome bridge function, we build intuition for the treatment bridge function: Eq.~\ref{PCIqu} specifies that for any particular $A=a$ and $X=x$, the effect of $U$ on the inverse probability of treatment assignment $A=a$ is equal to the effect of $U$ on some transformation of $Z$ ($q(Z,a,x)$). Note that for such a $q(Z,A,X)$ to exist, it is necessary that $Z$ to be associated with $U$ conditional on $A,X$ (except in the trivial case where $U=\emptyset$).\footnote{Assume for contradiction $Z\indep U\mid A, X$. Then $\mathbb{P}(A=a\mid U,X)^{-1}=\mathbb{E}[q(Z,a,X) \mid U,a,X]=\mathbb{E}[q(Z,a,X) \mid a,X]$. But if $U\not=\emptyset$, $\mathbb{P}(A=a\mid U,X)^{-1}$ must vary with $U$---a contradiction.} As with the outcome bridge function, $q(Z,a,X)$ is a function of the wrong variable ($Z$ rather than $U$), so for any fixed $(u,a,x)$ its value will vary with the realization of $Z$. However, Eq.~\ref{PCIqu} guarantees that this variation averages out: the conditional mean of $q(Z,a,X)$ given $U=u,A=a,X=x$ is exactly $\mathbb{P}(A=a\mid u,x)^{-1}$. In this sense, $q(Z,a,X)$ serves 
as a ``noisy" version of the unobservable inverse propensity score $\mathbb{P}(A=a\mid u,x)^{-1}$. Moreover, Eq.~\ref{PCIqu} characterizes $q(Z,A,X)$ as what the 
propensity score literature calls a \textit{valid weighting function}: despite depending on $Z$ rather than $U$, weighting outcomes by $q(Z,A,X)$ balances both observed and unobserved confounders, yielding an unbiased estimate of the average treatment effect \citep{nguyen_propensity_2020,lockwood_matching_2016}.



\subsubsection{Identification results of \citet{cui_semiparametric_2024}}
As in \citet{miao_identifying_2018,tchetgen_tchetgen_introduction_2024}, the formal identification results of \citet{cui_semiparametric_2024} begin from an observable integral equation (A.\ref{PCIq}, below) and use a completeness condition (A.\ref{compUW}) to show that any solution must also satisfy the $U$-level relationship in Eq.~\ref{PCIqu}. As before, the observable formulation provides the basis for estimation.\footnote{\citet{cui_semiparametric_2024} note in their Remark 6 that an alternative identification path (similar to \citet{miao_confounding_2024}) is to assume a bridge function satisfying Eq.~\ref{PCIqu} exists, which implies Eq. \ref{PCIq:eq}. Under an additional completeness condition (that $W$ is complete for $Z$ given $A,X$), A.\ref{compZW}, this yields the same identification result as their primary identification path.}

\begin{assumption}[Treatment Bridge Function]\label{PCIq}
For $a=0,1$ there exists a $q(Z,A,X)$ that solves 
\begin{align}\label{PCIq:eq} 
    \mathbb{P}(A=a\mid W,X)^{-1}
    =\mathbb{E}[q(Z,a,X)|W,A=a,X]
\end{align}
\end{assumption}

A new completeness condition is needed for identification:
\begin{assumption}[$W$ is complete for $U$ conditional on $A,X$]\label{compUW}
    $\mathbb{E}(g(U) \mid W, A,X) = 0 \iff g(U) = 0.$
\end{assumption}

Under this completeness condition (A.\ref{compUW}) and the conditional independence assumptions, the fact that $q$ averages ``correctly" for every value of $W$ (A.\ref{PCIq}) is sufficient to conclude that it 
also averages correctly for every value of $U$---that is, Eq.~\ref{PCIqu} holds.

Utilizing Eq. \ref{PCIqu} and base PCI assumptions, the potential outcome mean is identified \citep{cui_semiparametric_2024}:
\begin{align}
\mathbb{E}[Y(a)] = \mathbb{E}[I(A=a)Yq(Z,a,X)]
\end{align}

And thus the ATE is identified:\footnote{Many algebraic rearrangements are possible; \citet{cui_semiparametric_2024} uses the more compact form: $\tau_{q} = \mathbb{E}[(-1)^{(1-A)} Y q(Z, A, X)].$}
\begin{align}\label{eq_q_ID}
\tau_{q} =& \mathbb{E}[I(A=1)Y q(Z, 1, X)- I(A=0)Y q(Z, 0, X)]
\end{align}

This form closely parallels inverse probability weighting under unconfoundedness ($\tau {=} \mathbb{E}[AY e(X)^{-1}- (1-A) Y (1-e(X))^{-1}]$); with $q(Z,1,X)$ taking the place of $e(U,X)^{-1}$. 

As with the outcome bridge function identification results, an additional completeness condition (A.\ref{compWZ}) and regularity conditions (which can be found the supplement of \citet{cui_semiparametric_2024}) can be used to imply the existence of a treatment bridge function (A.\ref{PCIq}).

\addtocounter{assumption}{-2}
\begin{assumptionstar}[$Z$ is complete for $W$ conditional on $A,X$]\label{compWZ}
    $\mathbb{E}(g(W) \mid Z, A,X) = 0 \iff g(W) = 0.$
\end{assumptionstar}
\addtocounter{assumption}{1}

See the supplement of \citet{cui_semiparametric_2024} for detailed steps in their identification results.

\subsection{Completeness conditions}\label{sec_completeness_intro}

Four completeness conditions have been introduced across the two identification strategies, each with a distinct role but the same technical structure ($\mathbb{E}[g(\cdot) \mid *,a,x] = 0\ \iff g(\cdot) = 0.\  $). A.\ref{compUZ} and A.\ref{compUW} are what promote the observable integral equations (A.\ref{PCIh}, A.\ref{PCIq}) to the $U$-level properties (Eq.~\ref{PCIhu}, Eq.~\ref{PCIqu}) 
that are key to the identification results. A.\ref{compZW} and A.\ref{compWZ} play a different role: under regularity conditions \citep{miao_identifying_2018,cui_semiparametric_2024}, they imply the existence (and uniqueness) of the outcome and treatment bridge functions, respectively---though uniqueness of the bridge functions is not necessary for identification of the potential outcome mean \citep{miao_identifying_2018,cui_semiparametric_2024,zhang_proximal_2023}. Figure~\ref{fig:identification_paths} 
summarizes the two identification paths and highlights where each completeness condition enters. 

In this section, we discuss interpretation using A.\ref{compUZ} ($Z$ is complete for $U$ conditional on $A,X$) as a running example; the same interpretive logic applies to the other three conditions. When A.\ref{compUZ} was introduced in Section~\ref{sec_idh}, we 
described it as requiring ``sufficient variability'' of $Z$ 
relative to the variability of $U$ (as in \citet{tchetgen_tchetgen_introduction_2024}). \citet{andrews_examples_2017} provides a helpful basis for intuition on what this ``relative variability" means by formally connecting completeness to \textit{correlation}: for $Z$ to be complete for $U$ is equivalent to the condition that every non-constant, square-integrable function of $U$ is correlated with some square-integrable function of $Z$ (see \citet{andrews_examples_2017} Proposition 1 and Comment 4 for details). In the context of PCI identification via an outcome bridge function, this means that $Z$ carries enough information about $U$ that the bridge function (which depends on $U$ through the conditional distributions) can be recovered from observed data. 

We can make this more concrete through two practical questions:

\begin{enumerate}
    \item \textit{Relative dimensionality}: Does $Z$  have at least as many dimensions as $U$, and
    \item \textit{Component-wise relevance}: Is each dimension of $U$ reasonably thought to be associated with some component of $Z$ (given $A$ and $X$)?
\end{enumerate}

If the answer is a definitive ``no" to either of these questions, $Z$ is unlikely to be complete for $U$. 

While these are not formal sufficient conditions for completeness in the non-parametric form of A.\ref{compUZ}, we recommend focusing on them because: i) they may be more intuitive than the technical definition of completeness or the notion of ``sufficient" variability, ii) if they are not met, formal conditions (e.g., A.\ref{compUZ}) will not (or are unlikely to) be met, and iii) when these conditions are reasonably believed to hold, one may be able to invoke the arguments often made in the PCI literature that completeness conditions are commonly satisfied \citep{miao_identifying_2018,tchetgen_tchetgen_introduction_2024,cui_semiparametric_2024,miao_identifying_2023} (e.g., in exponential families \citep{newey_instrumental_2003,hu_nonparametric_2018}). The Appendix (Sections~\ref{ex:combo} and \ref{ex_comp_fail}) provides examples of parametric models where completeness does and does not hold, respectively. 

In the fully categorical case, these conditions can be 
stated precisely: relative dimensionality requires that within each combination of categories of $A$ and $X$, $Z$ has at least as many distinct categories as $U$, and completeness is then equivalent to the matrix with elements $\phi_{ij} =\mathbb{P}(U = u_i \mid Z = z_j, a, x)$ being full row rank \citep{miao_identifying_2023}.

We recommend Supplemental Section 2 of \citet{miao_identifying_2023} for additional background, examples, and citations to relevant literature related to completeness conditions, and \citet{andrews_examples_2017} for a deeper dive into completeness conditions in non-parametric settings. 

\begin{remark}\label{remark_udim_completeness}
    A practical difficulty is that $U$ is unobserved, so one must rely on substantive knowledge to determine the structure of confounding and how many ``dimensions'' $U$ has.\footnote{Even for completeness conditions relating the two sets of proxy variables (A.\ref{compZW}, A.\ref{compWZ}), $U$ is presumed to be the link between them, so the dimensionality of $U$ still matters.} There are several ways to think about the ``dimensionality" of $U$ or, to put it another way, the ways that $U$ meaningfully  varies. One is about whether $U$ is thought to be categorical, count, or continuous. For instance, confounding by education level might be adequately captured by a binary indicator for high school graduation, a categorical variable for highest degree, or a continuous measure of years of schooling---each implying different dimensionality for $U$, and thus different requirements for completeness. Another consideration is whether $U$ consists of multiple confounding mechanisms---such as ``health seeking" and ``physical health", which could be potentially related but distinct confounding mechanisms in a particular context. In this case, determining the dimensionality of $U$ is closely related to fundamental questions in latent factor analysis: how many latent factors are responsible for the correlation structure we observe \citep{kline_chapter_2016}? With these points in mind, it is important to develop a working ``conceptualization" of $U$ in order to consider whether proxies reasonably have sufficient dimensionality and association with the $U$, as well as to evaluate previously introduced assumptions that condition on $U$ (e.g., A.\ref{condZ}, A.\ref{condW}). 
\end{remark}


\begin{remark}
    Assumptions A.\ref{compUZ} and A.\ref{compUW}, that $Z$ and $W$ are complete for $U$, respectively, are fundamentally untestable given their reliance on the unobserved $U$. While the other two completeness conditions (A.\ref{compZW}, A.\ref{compWZ}) only involve observed data, they are also not empirically testable unless additional assumptions (such as linear structural models) are made \citep{canay_testability_2013}. However, there may be some testable \textit{implications} of completeness conditions if other assumptions (e.g., A.\ref{condZ} and A.\ref{condW}) hold also---see discussion in Section \ref{sec_empirical}.
\end{remark}

\subsection{Nonparametric Identification Summary}\label{sec_id_summary}

Utilizing the base assumptions required by most PCI methods (A.\ref{cons}, A.\ref{latpos}, A.\ref{latent}, A.\ref{condZ}, A.\ref{condW}), there are two main ``paths" to identification: via an outcome bridge function \citep{miao_identifying_2018,tchetgen_tchetgen_introduction_2024} and via a treatment bridge function \citep{cui_semiparametric_2024}. These bridge functions, respectively, serve as ``noisy" versions of the outcome mean $\mathbb{E}[Y\mid U,A,X]$ and inverse propensity score $\mathbb{P}(A=a\mid U,X)^{-1}$ and are the conceptual key to utilizing the two types of observed proxy variables to identification despite $U$ being unobserved (under formal assumptions provided). These two paths are summarized briefly here and in Figure \ref{fig:identification_paths}. 

First, under the additional assumptions that (i) an outcome bridge function exists (A.\ref{PCIh}, which can be implied via A.\ref{compZW} with regularity conditions \citep{miao_identifying_2018}) and (ii) $Z$ is complete for $U$ conditional on $A,X$ (A.\ref{compUZ}), Eq. \ref{PCIhu} holds and the ATE is identified as $\tau_h=\mathbb{E}[h(W,1,X)-h(W,0,X)]$ \citep{miao_identifying_2018}. 

Second, under the additional assumptions that (i) a treatment bridge function exists (A.\ref{PCIq}, which can be implied via A.\ref{compWZ} with regularity conditions \citep{cui_semiparametric_2024}) and (ii) $W$ is complete for $U$ conditional on $A,X$ (A.\ref{compUW}), Eq. \ref{PCIqu} holds and the ATE is identified as $\tau_q=\mathbb{E}[AY q(Z, 1, X)- (1-A) Y q(Z, 0, X)]$ \citep{cui_semiparametric_2024}. 

In the next section we discuss how the ATE can be estimated, building on these identification results. 

\section{Estimation of the Average Treatment Effect}\label{sec_estimation}

The identification results described in Sections ~\ref{sec_idh} and \ref{sec_idq} are \textit{nonparametric}, in the sense that they do not rely on parametric assumptions about the underlying data generating process. However, estimating the bridge functions in a fully nonparametric way can be challenging as these functions are defined as solutions to integral equations and may be unstable in practice. One common approach to address this difficulty is to impose parametric structure on the bridge function(s). In this section, we first discuss such methods that estimate the ATE using (i) an outcome bridge function with an assumed parametric form  \citep{tchetgen_tchetgen_introduction_2024,miao_confounding_2024,cui_semiparametric_2024}, (ii) the treatment bridge function with an assumed parametric form \citep{cui_semiparametric_2024}, (iii) both bridge functions with specified parametric forms, used to form doubly robust estimators \citep{cui_semiparametric_2024}. We then briefly discuss (iv) nonparametric estimation of the ATE which typically requires some form of regularization mitigate instability in the estimation of the bridge functions \citep{ghassami_minimax_2021,kallus_causal_2022,mastouri_proximal_2023}. 

\subsection{Estimation via a Parametric Outcome Bridge Function}\label{sec_h_est}
We begin with discussion of methods for estimating the ATE using the proximal g-formula ($\tau_h = \mathbb{E}[h(W,1,X)-h(W,0,X)]$), assuming a specific parametric form for the outcome bridge function. 
\begin{assumption}\textbf{Parametric Outcome Bridge Function}\label{para_h}
    The outcome bridge function has a specific parametric form $h(W,A,X;\eta)$, where $\eta$ is a finite dimensional, real-number-valued parameter. 
\end{assumption}

This is a stronger version of A.\ref{PCIh} (which only assumes that an $h(W,A,X)$ solving Eq. \ref{PCIh:eq} exists). For now, we set aside the question of how to justify the choice of a particular form of $h(W,A,X;\eta)$--- this is returned to in Section \ref{sec_closed_form}. 
  The most common form assumed in the PCI literature is linear, such as:
    \begin{equation}\label{eq:linear_h}
        h(W,A,X;\eta)=\eta_0+\eta_wW+\eta_aA +\eta_xX
    \end{equation}
Or, more compactly, 
$h(w,a,x;\eta)=\vecWAX\eta$, where $\eta=(\eta_0,\eta_w,\eta_a,\eta_x)^T$.

Estimation of $\eta$ then yields the following estimator of the ATE: $\hat\tau_h=P_n[h(W,1,X;\hat\eta)-h(W,0,X;\hat\eta)]$, where the $P_n$ denotes taking the sample average. In the simple linear case (Eq. \ref{eq:linear_h}), $\tau_h=\eta_a$ is implied. Our challenge is thus to estimate $\eta$. 

\subsubsection{Estimating Equations for Bridge Function Parameters}\label{sec_h_ee}

The definition of an outcome bridge function (Eq. ~\ref{PCIh:eq} in A.\ref{PCIh}) can be rearranged as:
$$\mathbb{E}[Y - h(W,A,X;\eta)\mid Z,A,X]=0.$$

This motivates estimating equations such as \citep{miao_confounding_2024,cui_semiparametric_2024}: 
\begin{equation}\label{eq:pci_h_ee}
    P_n[(Y-h(W,A,X;\eta))(1,Z,A,X)^T]=0
\end{equation}

It may be possible to solve this system of equations directly to obtain an estimate $\hat\eta$ that satisfies Eq.~\ref{eq:pci_h_ee}, or to utilize generalized method of moments (GMM) \citep{harris_introduction_1999,stefanski_calculus_2002}. This is the approach suggested in \citet{miao_confounding_2024}. These estimation methods also specify how standard errors can be estimated for the $\hat\eta$, and thus, standard errors for $\hat\tau_h$ can be obtained via the delta method \cite{stefanski_calculus_2002,harris_introduction_1999} or by setting up an additional component of the estimating equations for the proximal g-computation step \citet{miao_confounding_2024}. Further details on this estimation approach can be found in Appendix \ref{appendix_estimation}.

 \begin{example}
    Consider the simple linear case where we assume $h(W,A,X;\eta)=\vecWAX\eta$ and we want to solve estimating equations:
     $$P_n[(Y-\vecWAX\eta)(1,Z,A,X)^T]=0$$ 
  There are as many unknowns (dim($\eta$)=4) as equations (4) so this system of equations can be solved directly, as long as the first matrix in the rearrangement below is invertible. 
    $$P_n\begin{bmatrix}
        1& W&A&X\\
        Z&ZW&ZA&ZX\\
        A&AW&A^2&AX\\
        X&XW&XA&X^2
    \end{bmatrix}\begin{bmatrix}
        \eta_0\\\eta_w\\\eta_a\\\eta_x
    \end{bmatrix}=P_n\begin{bmatrix}
        Y\\YZ\\YA\\YX
    \end{bmatrix}$$
\end{example}

\begin{remark}\label{remark_ee_ols}
    If the form of estimating equations like Eq. ~\ref{eq:pci_h_ee} is unfamiliar, it may help to note parallels to the conditions that define a standard outcome regression. For example, the regression of $Y$ on some set of features $\tilde X$, assuming a form $\mathbb{E}[Y\mid \tilde X;\beta]=\mu(\tilde X;\beta)$, aims to estimate the true value of $\beta$ that solves: 
$$\mathbb{E}[Y - \mu(\tilde X;\beta)\mid\tilde X]=0.$$

This motivates the estimator $\hat\beta$ that is the solution to the estimating equations: 
$$P_n[(Y - \mu(\tilde X;\hat\beta))\tilde X^T]=0.$$
This is simply another way to write conditions of standard  regression. For example, if $\mu(\tilde X;\hat\beta)=\tilde X\beta$, this is the same as setting the derivative of the least squares criteria equal to zero.
 
The PCI estimating equations (Eq.~\ref{eq:pci_h_ee}) look similar to this standard case (with $h$ representing a conditional mean instead of $\mu$). However, a key difference and additional complexity is that $h$ is a function of $W$, while the conditioning in Eq.~\ref{PCIh:eq} involves $Z$, and not $W$. While this variation means that the estimating equations given by Eq. \ref{eq:pci_h_ee} are not equivalent to a simple one-stage regression, standard methods \citep{stefanski_calculus_2002,harris_introduction_1999} still exist for estimation. 
\end{remark}

\begin{remark}\label{remark_efficiency}
    While a full exploration of the efficiency of PCI estimators is out of our scope, we note that \citet{cui_semiparametric_2024} derives the form of the efficient influence function for the parameters of the outcome bridge function which can be used to determine the most efficient form of estimating equations for the outcome bridge function parameters. However, they note that operationalizing this result would require making additional assumptions on the observed data distribution ($\mathbb{P}(Y,W\mid Z,A,X)$) beyond the parametric form of $h(W,A,X)$, and that in practice, the efficiency gains made by utilizing these additional assumptions may be modest \citep{cui_semiparametric_2024}. Thus, when estimating the ATE via outcome bridge function in simulations and an applied example, \citet{cui_semiparametric_2024} forgoes such optimization and uses estimating equations in the form of Eq. \ref{eq:pci_h_ee}. \citet{cui_semiparametric_2024} similarly derives the form of the efficient influence function for the treatment bridge function parameters, where the same caveats apply. 
\end{remark}
 
\subsubsection{Two-Stage Estimation of the ATE}\label{sec_proximal_g_comp}

In some specific scenarios, PCI estimation can be reduced to a two-stage regression procedure \citep{tchetgen_tchetgen_introduction_2024}. For example, assuming $$h(W, A, X; \eta) = \eta_0 + \eta_aA + \eta_xX + \eta_w W,$$ and plugging this into Eq. \ref{PCIh:eq} the conditional mean of $Y$ is:
\begin{align}
\label{eq:two_stage_mean}
    \mathbb{E}[Y \mid& Z, A, X; \eta]= \eta_0 + \eta_a A + \eta_x X+ \eta_w \mathbb{E}[W \mid A, Z, X]
\end{align}

This motivates a two-stage estimation approach:
\begin{enumerate}
    \item Estimate $\widehat{W} = \hat{\mathbb{E}}[W \mid A, Z, X]$ (e.g., via linear regression of $W$ on $Z,A,X$). 
    \item Regress $Y$ on $(\widehat{W}, A, X)$ using linear regression.
\end{enumerate}
Then, the estimated ATE is $\hat{\eta}_a$. 

The same approach can also be motivated by assuming linear models conditional on $U$ \citep{tchetgen_tchetgen_introduction_2024,liu_regression-based_2024,zivich_introducing_2023}: 
\begin{align*}
    \mathbb{E}[Y \mid A, Z, X, U] &= \gamma_0 + \gamma_a A + \gamma_u U + \gamma_x X, 
\mathbb{E}[W \mid A, Z, X, U] = \omega_0 + \omega_u U + \omega_x X,
\end{align*}
which also imply Eq. \ref{eq:two_stage_mean} (and thus, that a linear form of $h(W, A, X; \eta)$ would satisfy Eq.~\ref{PCIh:eq})---see Appendix Section \ref{appendix_very_special}.

A further special case is estimating $\hat{\mathbb{E}}[W \mid A, Z, X]$ via linear regression (e.g, assuming ${\mathbb{E}}[W \mid A, Z, X;\theta]= \theta_0 + \theta_aA +\theta_x X + \theta_zZ$)---this ``very special case" is exactly equivalent to solving the system of equations (Eq.~\ref{eq:pci_h_ee}) in the example in the prior section. Further explanation is provided in Appendix Section~\ref{appendix_very_special}.

Considering this type of two-stage regression procedure can be motivated in two ways. The first is simplification: framing estimation as two-stage regression may be simpler to understand and implement than a more general estimating equation based approach---this is particularly true of the two-stage linear regression, the ``very special case." The second reason, though not discussed explicitly in \citet{tchetgen_tchetgen_introduction_2024}, is efficiency: as noted in the previous section, the form of the efficient influence function relies on assumptions about additional components of the distribution \citep{cui_semiparametric_2024}---as does a two stage approach. 
The generalization of this approach (beyond linear $h(W,A,X;\eta)$) as described in \citet{tchetgen_tchetgen_introduction_2024} is discussed in Appendix Section~\ref{appendix_g_comp}.

\subsection{Estimation via a Parametric Treatment Bridge Function}\label{sec_q_est}

Consider now that instead of assuming a parametric form for the outcome bridge function, we are willing to assume a parametric form of the treatment bridge function and wish to use the identification results of \citet{cui_semiparametric_2024}: $\tau_q=\mathbb{E}[AY q(Z, 1, X)]- \mathbb{E}[(1-A)Y q(Z, 0, X)]$.
\begin{assumption}\textbf{Parametric Treatment Bridge Function}
    The treatment bridge function has a specific parametric form $q(Z,A,X;\phi)$ where $\phi$ is a finite dimensional real number-valued parameter. 
\end{assumption}

A simple example of a parametric form for the treatment bridge function, as used in \citet{cui_semiparametric_2024} is:
$$ q(Z,A,X; \phi) = 1 +\exp\left[(-1)^{1-A}((1,Z,A,X)\phi)\right].$$
If $f(Z|W,A,X)$ normally distributed, this $q(Z,A,X; \phi)$ implies a logistic form for $\mathbb{P}(A=a\mid W,X)$ \citep{cui_semiparametric_2024}. 

The defining equation for the treatment bridge function (Eq.~\ref{PCIq:eq}) motivates estimating equations, as previously discussed for the outcome bridge function. First, with some rearrangement (see Appendix~\ref{ee_q_derivation}) Eq. \ref{PCIq:eq} is equivalent to: 

\begin{equation}
    \mathbb{E}\!\left[
\begin{pmatrix}
(-1)^{1-A}q(Z,A,X)A\\
(-1)^{1-A}q(Z,A,X)
\end{pmatrix}
-
\begin{pmatrix}
1\\
0
\end{pmatrix}
\Bigg| W,X
\right]
=
\begin{pmatrix}
0\\
0
\end{pmatrix}.
\end{equation}

Estimating equations for the treatment bridge function can then take the following form, as used in \citet{cui_semiparametric_2024} (see Appendix~\ref{ee_q_derivation} for further detail): 
\begin{align}\label{eq:pci_q_ee}
P_n[
(-1)^{1-A} & q(Z,A,X;\hat\phi)(1,W,A,X)^T\\
&-(0,0,1,0)^T
]=
(0,0,0,0)^T\notag
\end{align}

This can be estimated similarly to the outcome bridge functions (e.g., by GMM) \citep{cui_semiparametric_2024}. 

\begin{remark}
    As with the outcome bridge function, \citet{cui_semiparametric_2024} derive the form of the efficient influence function for parameters of the treatment bridge function. However, the same caveats apply here: i) finding the form of the efficient influence function may require additional assumptions, and ii) the efficiency gains may not be worth the additional trouble and assumptions in practice \citep{cui_semiparametric_2024}. 
\end{remark}

\subsection{Semiparametric Doubly Robust ATE Estimation}\label{sec_DR_est}

While the outcome and treatment bridge identification results can be used separately, \citet{cui_semiparametric_2024} show how use of both bridge functions can yield an estimator that is robust to misspecification of one of the two parametric bridge functions---\textit{doubly robust}. 

This estimator takes the following form (which is based on the efficient influence function for the ATE) \citep{cui_semiparametric_2024}:
\begin{align}\label{tau_dr}
    \hat\tau_{dr} = \mathbb{P}_n
\Bigl[(-1)&^{1-A}\,q(Z,A,X;\hat \phi)\,(Y - h(W,A,X;\hat\eta))\notag
  \;\\&+\;h(W,1,X;\hat\eta) - h(W,0,X;\hat\eta)\Bigr].
\end{align}

\citet{cui_semiparametric_2024} suggest estimating the bridge function parameters using the estimating equations of the forms introduced in the previous sections (Eqs. \ref{eq:pci_h_ee} \& \ref{eq:pci_q_ee}), using GMM. The estimated bridge functions are then plugged into equation~\ref{tau_dr}. For inference, standard errors can be obtained by estimating the variance of the efficient influence function \cite{cui_semiparametric_2024}. 

This estimator is \textit{consistent} (asymptotically unbiased) and asymptotically normal (under regularity conditions \citep{cui_semiparametric_2024}) if at least one of the parametric bridge function models ($h(W,A,X; \eta)$ or $q(Z,A,X; \phi)$) is correctly specified---this is the \textit{double robustness} property in a PCI context. This also implicitly requires that the identification assumptions sufficient to guarantee existence of the corresponding bridge function are satisfied.  When both models ($h(W,A,X; \eta)$ and $q(Z,A,X; \phi)$) are correctly specified, bridge functions are unique (which is implied if A.\ref{compWZ} and A.\ref{compZW} hold), and additional regularity conditions are met (see details in \citep{cui_semiparametric_2024}), the estimator can achieve the semiparametric efficiency bound (i.e., the variance of the efficient influence function)\citep{cui_semiparametric_2024}.

\begin{remark}\label{remark_ATT}
    In addition to their ATE results, \citet{cui_semiparametric_2024} provide a semiparametric PCI approach for the average treatment effect on the treated (ATT; $\mathbb{E}[Y(1)-Y(0)|A=1]$), which is often policy-relevant. As ATT identification requires identification of $\mathbb{E}[Y(0)\mid A=1]$ but does not require $\mathbb{E}[Y(1)\mid A=0]$ (and since $\mathbb{E}[Y(1)\mid A=1]$ is observed under the assumption of consistency A.\ref{cons}), the conditions required are slightly weaker than ATE estimation. Alternate forms of the bridge functions, regularity conditions, and efficient influence function are given in Appendix G of \citet{cui_semiparametric_2024}.
\end{remark}

\subsection{Specifying Parametric Forms of Bridge Functions}\label{sec_closed_form}
As noted in Remark~\ref{remark_known_forms}, specifying parametric bridge functions avoids some of the complexity of solving integral equations in nonparametric settings. In addition to the linear outcome bridge function ($h(W,A,X)=\vecWAX\eta$), additional examples of such parametric assumptions for the outcome bridge function include a logistic form ($\logit(h(W,A,X))=\vecWAX\eta$) when $Y$ is binary 
\citep{tchetgen_tchetgen_introduction_2024} or a log link for count data ($\log(h(W,A,X))=\vecWAX\eta$) \citep{miao_confounding_2024}. For the treatment confounding bridge functions an additional option is to use a linear form of $q(Z,A,X;\phi)$--- even with a binary treatment assignment, this may be reasonable as an approximation. Additionally, the bridge functions may include interactions between variables; \citet{cui_semiparametric_2024} include interaction terms between $A$ and $Z$, and $Z$ and $X$ in their treatment bridge function. In principle, parametric bridge functions could be made to be quite complex (e.g., adding some spline terms \citep{cobzaru_bias_2022}), though at a certain point overfitting would become a concern. This motivates the discussion of nonparametric estimation methods in Section~\ref{sec_np_est}.  

A reasonable starting point, given the interpretation of the bridge functions described in Sections~\ref{sec_idh} and~\ref{sec_idq}, is to consider what parametric forms would be reasonable for $\mathbb{E}[Y|Z,A,X]$  and  $\mathbb{P}(A=a|W,X)^{-1}$ and specify that $h(W,A,X;\eta)$ and $q(Z,A,X;\phi)$ match these forms.\footnote{One could also consider $\mathbb{E}[Y|U,A,X]$  and  $\mathbb{P}(A=a|U,X)^{-1}$, in the vein of \citet{miao_confounding_2024}.} The technical caveat is that when the outcome bridge function (or treatment bridge function) is not linear, its form is not only dependent on $\mathbb{E}[Y|Z,A,X]$ ($\mathbb{P}(A=a|W,X)^{-1}$) but on $f(W|Z,A,X)$ ($f(Z|W,A,X)$)---which captures how the proxies related to each other conditional on $A,X$, through their relationships with $U$. If the relationship between $U$ and $Z$ (or $U$ and $W$) is non-linear, this could create complications. While $q(Z,A,X; \phi) = 1 +\exp\left[(-1)^{1-A}((1,Z,A,X)\phi)\right]$ corresponds to a logistic form of $\mathbb{P}(A=a|W,X)^{-1}$ when $f(Z|W,A,X)$ is a normal distribution, if we wanted to assume a different $f(Z|W,A,X)$ (say, if proxies were binary), the implied form of $\mathbb{P}(A=a|W,X)^{-1}$ may not be simple. This type of technicality underscores why simple, normally distributed proxies, linear structural models, or fully categorical data are often used in illustrative examples in the PCI literature.

Difficulty may arise in scenarios where the data generating model is known, but deviates from the special cases often used in the PCI literature.  
For example, the simulations of \citet{cui_semiparametric_2024} are conducted with a data generated from a model where $\mathbb{P}(Y,Z,W,U\mid A,X)$ is multivariate normal and $\mathbb{P}(A \mid X)$ follows a logistic model \citep{cui_semiparametric_2024}. In this case, the simple closed-form expressions for $h(W,A,X;\eta)$ and $q(Z,A,X;\phi)$ given as examples in the previous sections can be derived, and $\eta$ and $\phi$ could in principle be written in terms of parameters from the data generating model \citep{cui_semiparametric_2024}. However, if we consider a slight variation of the normal and logistic data distribution, things are not so clean. Assume that we want to simulate data where $\mathbb{P}(Z,W,U\mid X)$ follows a multivariate normal distribution and $A$ follows a logistic model conditional on $Z,W,U,X$, ($A \mid U,Z,W,X \sim \text{Bernoulli}(\text{expit}(\alpha_0 + \alpha_u U + \alpha_x X + \alpha_z Z))$), and $Y$ is normally distributed conditional on $A,W,Z,U,X$. This data generating model is a natural choice in many simulation settings given that proxies may often be measured pre-treatment.\footnote{Such a model is used in \citet{nguyen_propensity_2020} and \citet{hong_bayesian_2017}.} The difference from the previous example is subtle: the distribution of $(Z,W,U)$ is defined conditional on $X$ and not conditional on $A$, and the distribution of $A$ is defined conditional on $U,W,Z$ and $X$. In this DGM, the implied $\mathbb{P}(Z \mid W,A,X)$ and $\mathbb{P}(W \mid Z,A,X)$ are not gaussian and integral equations for $h(W,A,X;\eta)$ and $q(Z,A,X;\phi)$ do not admit simple closed-form solutions.

Still, when a closed form bridge function is unavailable, or in real-world applications where the data generating model is unknown and likely complex, using a closed form like the examples provided above is a practical starting point. \citet{miao_confounding_2024} suggest that to probe how sensitive a PCI estimate is to the form of its parametric bridge function assumption, one could also estimate the ATE with nonparametric techniques or more complex parametric models, and compare the estimates. Comparing the performance of PCI estimators under various levels of parametric misspecification, and investigation of practical tradeoffs in bias and variance with semi-parametric and nonparametric PCI estimators (discussed in the next section) is an area in which additional guidance could aid practitioners.

\subsection{Nonparametric estimation}\label{sec_np_est}
Nonparametric estimation strategies avoid parametric assumptions about the form of the bridge functions or other parts of the data distribution. In principle, these methods can approximate arbitrarily complex relationships among treatment, covariates, and proxies. For example, when all variables are binary or categorical, nonparametric estimation of the bridge functions is equivalent to fitting saturated models \citep{shi_multiply_2020} (see Appendix~\ref{ex_binary_np}). However, as the set of covariates and proxies grows, so does the complexity of the function--- typically requiring more data and making inference more challenging. 

Most nonparametric PCI estimation approaches can be seen as generalizations of the estimating approaches described previously. They utilize the same definitions of the bridge functions, but instead of assuming $h(W,A,X)$ or $q(Z,A,X)$ lie in a finite-dimensional parametric families, these methods search for solutions in richer function spaces. For example, \citet{mastouri_proximal_2023} introduce kernel-based method of moments and two-stage approaches that parallel the approaches of estimating equation approach \citep{miao_confounding_2024} and two-stage regression \citep{tchetgen_tchetgen_introduction_2024} approaches described in the previous sections. 

A key work in this space is \citet{ghassami_minimax_2021}, which presents estimators that solve a minimax optimization procedure. \citet{ghassami_minimax_2021} provide convergence results in the case of a reproducing kernel Hilbert space, where regularization is used.\footnote{This regularization addresses the potential ill-posedness of the integral equations.} Thus, implementation requires choosing kernels and tuning parameters (e.g., bandwidths and regularization coefficients), typically selected via cross-validation \citep{chernozhukov_doubledebiased_2024}. \citet{ghassami_minimax_2021} show that these estimators can achieve a form of double robustness related to convergence rates of the nuisance functions involved in the minimax problem. \citet{kallus_causal_2022} also proposes a similar flexible nonparametric estimation framework, which can accommodate kernel methods or other function classes such as neural networks. Both \citet{kallus_causal_2022} and \citet{ghassami_minimax_2021} provide in-depth summaries of how their work relates to the broader literature. Another notable work in this space is \citet{zhang_proximal_2023}, which proposes sieve-based estimators to explicitly address estimation in scenarios where bridge functions are not-unique. 

\subsection{Conclusion: Estimation}\label{sec_est_summary}
Selection of approaches for estimation of causal effects under PCI involves trade-offs. Methods that assume parametric forms of bridge functions may be simpler to implement than nonparametric methods, particularly in special cases when they correspond to a two-stage regression estimator \citep{tchetgen_tchetgen_introduction_2024}. Doubly robust semi-parametric estimators \citep{cui_semiparametric_2024} offer protection against misspecification of one bridge function model, but still require the other model to be properly specified for consistency to hold. Nonparametric methods are more flexible but require more attention to convergence rates and may be more complex to implement (i.e., require cross fitting, penalization, choice of kernel and hyperparameters) \citep{ghassami_minimax_2021}. An additional consideration is that while some PCI papers make code available, more general and accessible implementations (such as R-packages) are not widely available at this time, except for special cases like two-stage regression \citep{pci2s_package}.

\section{Extensions in PCI Identification and Estimation}\label{sec_extensions}

Our discussion of identification (Sections~\ref{sec_idh} and \ref{sec_idq}) and estimation  (Section~\ref{sec_estimation}) focused on the ATE ($\mathbb{E}[Y(1)-Y(0)]$) with binary treatments and single time-point outcomes. However, PCI has been extended to a variety of additional scenarios \citep{ying_proximal_2021,sverdrup_proximal_2023,li_doubly_2022,wu_doubly_2023,park_proximal_2025,liu_proximal_2023,chen_proximal_2024,ghassami_causal_2023, zhou_causal_2024}. As a full review of all PCI-related work is out of our scope, this section discusses two particular extensions that are closely related to the settings in Section~\ref{sec_estimation} and practical guidance on proxy selection to come in Section~\ref{sec_practical}: (i) the two-stage regression-based approach of \citet{liu_regression-based_2024}, which targets estimands corresponding to regression coefficients in generalized linear models (generally different than the standard ATE) and (ii) emerging methods for quantifying and correcting bias when some PCI assumptions are invalid \citep{ghassami_partial_2023,huang_relative_2025,cobzaru_bias_2022,yu_fortified_2025,rakshit_adaptive_2025}. 

\subsection{Two-stage Generalized Linear Models}\label{sec_reg_based}

\citet{liu_regression-based_2024} propose an estimating approach that utilizes two generalized linear models and aims to estimate the coefficient on $A$ in a regression second stage regression for outcome $Y$.  In the case of continuous $Y$, the coefficient on $A$ in the second stage model of \citet{liu_regression-based_2024}'s approach does correspond to the ATE ($\mathbb{E}[Y(1)-Y(0)]$) and yields the exact same estimation procedure of two stage linear regression as for the simple linear bridge function described in \citet{tchetgen_tchetgen_introduction_2024}. However, the identification and estimation approaches of \citet{liu_regression-based_2024} and \citet{tchetgen_tchetgen_introduction_2024} are not generally the same.

\citet{liu_regression-based_2024} assume parametric models for conditional means of $Y$ and $W$ given the unobserved confounder $U$ and other parts of the observed data. The form of these assumptions and the estimation procedure vary based on the data type of the outcome and proxies (continuous, count, or categorical)---see Appendix Figure 2 in \citet{liu_regression-based_2024}. 

The estimand in each case is a coefficient on $A$ in the assumed model for $Y$ \citep{liu_regression-based_2024}. For example, when $Y$ is binary, the estimand is a conditional log odds ratio: 
$$\beta_a=\log\left(\frac{\frac{P(Y=1\mid U, A=1, X)}{P(Y=0\mid U, A=1, X)}}{\frac{P(Y=1\mid U, A=0, X)}{P(Y=0\mid U, A=0, X)}}\right)$$

Note both the scale (multiplicative) and conditioning (on $U,X$). 

Another key difference from the two-stage approach of \citet{tchetgen_tchetgen_introduction_2024} is that bridge functions are not used for identification in \citet{liu_regression-based_2024}: justification for their procedure is provided case-by-case in their Appendix A \citep{liu_regression-based_2024}. The estimators in \citet{liu_regression-based_2024} generally also do not have the theoretical robustness to some parametric misspecification that doubly-robust semi-parametric PCI estimators do \citep{cui_semiparametric_2024}. However, an advantage of this approach is that an R-package (\textit{pci2s}) \citep{pci2s_package} is available for implementation.

\subsection{PCI With Invalid or Incomplete Proxies}\label{sec_invalid}

The focus of this work has been on the standard PCI framework, which assumes valid proxies and that completeness conditions are satisfied. However, recent work has begun to address settings in which some of these assumptions may fail. Assuming that at least a known fraction of treatment confounding proxies are valid, \citet{yu_fortified_2025} introduce a semiparametric, influence-function-based estimator that parallels the approach of \citet{cui_semiparametric_2024}, with additional assumptions and estimation steps to account for potential invalid proxies. Similarly, \citet{rakshit_adaptive_2025} consider a setting in which up to 50\% of treatment confounding proxies and 50\% of outcome confounding proxies may be invalid, proposing an adaptive LASSO-based procedure within linear structural models that simultaneously selects valid proxies and estimates the treatment effect. Additionally, with invalid or incomplete proxies or even when only one type of proxy is available, it may still be possible to obtain informative bounds on the ATE (rather than point identification) \citep{ghassami_partial_2023} or to identify a different estimand \citep{park_single_2024}. 

A complementary line of work focuses on bias quantification under violated PCI assumptions. \citet{cobzaru_bias_2022} provide analytic derivations of bias when completeness conditions fail, within a linear structural model that includes interactions between $A$ and $U$ and excludes direct effects of $W$ on $Y$. Extending this perspective, \citet{huang_relative_2025} propose a framework for assessing bias of PCI and IV two-stage estimators (assuming linear structural models) relative to regression-adjustment estimators that assume unconfoundedness, in scenarios where identification assumptions are not perfectly satisfied. Collectively, this emerging literature highlights promising directions for extending PCI methods to more realistic settings where PCI assumptions are not guaranteed to hold.

\section{Practical Considerations in Operationalizing PCI}\label{sec_practical}
In this section, we outline key considerations for proxy selection in applied PCI analyses.
We assume the researcher has a specific conceptualization of the unobserved confounding mechanisms $U$ for their example unless otherwise noted, and that the other elements important to any causal analysis such as the target population, treatment, outcome, time-frame, estimand, and data source are also carefully defined \citep{hernan_target_2022}. 

\subsection{Selecting Proxy Variables}\label{sec_selection}
Proxy selection relies first and foremost on substantive knowledge to identify variables that are reasonably believed to satisfy PCI assumptions for the unobserved confounder(s) of concern. In this section, we discuss 1) how variables that would traditionally be seen as \textit{error-prone measurements} of $U$ can serve as proxies 2) how traditional \textit{negative control variables} may be used as proxies, and 3) how \textit{temporal reasoning} can helps rule out causal relationships, with some cautionary notes. 

\subsubsection{Error-prone measurements}\label{sec_measurement}
Measurement error refers to the discrepancy between the true value of a variable and the value as measured or observed in the data \citep{innes_measurement_2021}. In some cases, it may be natural to think of $U$ as a latent factor or a set of latent factors that are measured by a collection of indicators, and these indicators could be used to form our sets of proxies. A PCI example with a measurement error perspective is estimation of the effect of right heart catheterization in the ICU on mortality, where biomarkers derived from blood tests are viewed as error-prone measurements of underlying ``physiological state" \citep{tchetgen_tchetgen_introduction_2024,liu_regression-based_2024,cui_semiparametric_2024}. 

Two important considerations in this setting where we have measurements of an unobserved confounder are i) whether unobserved confounding is truly a concern given that we observe the measurements and, if so, ii) whether a particular measurement can in fact be used as a treatment confounding proxy, outcome confounding proxy, either, or neither. First, if all relevant information for the treatment decision is captured in observed variables, controlling for these could be sufficient to control for confounding. Imagine a treatment assignment was made by a physician purely by reviewing lab results and other recorded clinical history: controlling for these could be sufficient to satisfy unconfoundedness, even if these variables are not a perfect picture of patient health. However, in many cases, treatment assignment is likely informed by clinician \textit{perception} of patient health status and is also informed by patient preferences, which may not be perfectly captured in observed data---then unobserved confounding may still be a problem and use of PCI may still be motivated. 

Second, it should be noted that measurements that are believed to be used directly in treatment assignment (i.e. the lab results in the last example) may be reasonable treatment confounding proxies (if they are not linked to the outcome other than through $U, A, X$), but are less likely to be reasonable outcome confounding proxies, given the direct link to treatment. However, if there are observed measurements that are \textit{known} to be unused in treatment assignment, these could be good candidate outcome confounding proxies. For example, if participants in a study complete self-reported health surveys that are by design not available to a physician involved in the treatment decision, the survey data may provide good candidates for outcome confounding proxies \citep{tchetgen_tchetgen_introduction_2024}. As the participant generally also has influence over treatment choice, this argument would also require that it is reasonable to believe the survey response is only linked to treatment assignment through the unobserved confounder (or observed factors).
Another example would be features derived from unstructured text (e.g., embeddings derived from clinical notes)---these features can be seen as capturing the latent information in the notes, and the embeddings themselves could be assumed to not play a role in the treatment assignment mechanism or to directly influence health outcomes, aside from their relationship to the ``true confounding" information contained in the text \cite{chen_proximal_2024}. 

\subsubsection{Negative controls.}\label{sec_NC} Negative control exposures share the same conditional independence structure (A.\ref{condZ}) as treatment confounding proxies, while negative control outcomes share the same structure as outcome confounding proxies (A.\ref{condW}) \citep{shi_selective_2020}. The terms ``negative controls" and ``proxies" are effectively interchangeable within the PCI context \citep{shi_multiply_2020,tchetgen_tchetgen_introduction_2024,xie_automating_2024}. 

However, there are differences between how the term ``negative control" is used in broader contexts, relative to its use in PCI. First, negative control exposures are often specifically other ``exposures" that are believed to be influenced by the same unobserved confounders as the ``primary" exposure ($A$), but are believed to not affect the primary outcome ($Y$) \citep{lipsitch_negative_2010}. Negative control outcomes are often other ``outcomes" that are believed to be influenced by the same unobserved confounders as the ``primary" outcome ($Y$) but are believed not to be affected by the primary exposure ($A$). Thus, this context differs a bit from the measurement error perspective---it might not feel natural to call a negative control a ``measurement" of an unobserved confounder. This broader scope is why the more general term ``proxy" is proposed in \citet{tchetgen_tchetgen_introduction_2024}.

Second, negative controls are often used for ``bias detection" rather than for identification (also known as ``bias correction") \citep{shi_selective_2020,yang_advances_2024}. Bias \textit{detection} methods may only require one type of proxy and utilize different assumptions. For example, they may not assume any completeness conditions, though they do generally require $U$-relevance (often called ``$U$-comparability" when utilizing the terminology of negative controls \citep{shi_selective_2020}). 

Reviews of negative control use such as \citet{zafari_state_2024} or \citet{shi_selective_2020} may be helpful resources for PCI users \citep{tchetgen_tchetgen_introduction_2024}. However, many negative control examples do not use both a negative control exposure \textit{and} a negative control outcome\footnote{While some negative control methods for bias correction are designed to use only one type of negative control, they typically require additional assumptions--- see \citet{yang_advances_2024} and references within for details.}--- and in PCI, both types of proxies are needed, except in some special cases such as for partial identification \citep{ghassami_partial_2023}. An example of proximal causal inference using proxies that may be traditionally viewed as negative controls is \citet{shi_multiply_2020}---see the description in Section \ref{sec_pci} of their vaccine safety application. 

Considering either error-prone measurements or negative controls as potential proxies does not change the set of assumptions to be satisfied, but it does provide somewhat of a different perspective on the type of variable we would want to use as a proxy. Approaching the task of finding proxies from both these perspectives may yield a richer understanding of the possible proxies and provide multiple entry points for applied researchers.  

\subsubsection{Temporal Reasoning.}\label{sec_temporal} An additional tool in identifying proxies is the \textit{arrow of time:} the future cannot cause the past. Pre-treatment measurements are sometimes suggested as candidate outcome confounding proxies, since they cannot be caused by subsequent treatment assignment \citep{tchetgen_tchetgen_introduction_2024, cui_semiparametric_2024, liu_regression-based_2024, shi_selective_2020}. However, pre-treatment measures should only be used as outcome confounding proxies if it is reasonable to assume that they are not directly involved in treatment assignment---see discussion in Section~\ref{sec_measurement}. Similarly, while post-outcome variables have been proposed as a source for candidate treatment confounding proxies \citep{tchetgen_tchetgen_introduction_2024, cui_semiparametric_2024, liu_regression-based_2024}, their validity depends critically on the outcome not directly influencing them, which can be a strong assumption in many settings. 

\subsection{Completeness assumptions}\label{sec_practical_completeness}

In addition to the conditional independence assumptions (A.\ref{condZ} and A.\ref{condW}), treatment confounding proxies and outcome confounding proxies should also plausibly satisfy the requisite completeness conditions for the identification approach used, else the PCI estimator may be biased \citep{cobzaru_bias_2022}. However, assessing whether formal completeness conditions hold in a particular applied setting may be challenging, as discussed in Section~\ref{sec_completeness_intro}. It is sensible to focus on the following criteria, which should be satisfied for either identification path \citep{miao_identifying_2018,cui_semiparametric_2024}:
\begin{enumerate}
   \item All proxy variables (in both $Z$ and $W$) should be believed to be $U$-relevant (i.e., associated with some component of $U$ conditional on $A$ and $X$).
   \item The set of treatment confounding proxies ($Z$) should capture at least as many meaningful dimensions as are believed to exist in $U$, as should the set of outcome confounding proxies ($W$).
\end{enumerate}

While it may be difficult in practice to determine how many dimensions $U$ has (see Remark~\ref{remark_udim_completeness}), one can still look out for cases in which $Z$ and $W$ clearly have lower dimensionality than $U$. For instance, completeness would be suspect if $U$ is thought to be highly complex---consisting of distinct confounding mechanisms such as health-seeking and access as well as underlying health status---but we only have access to one binary treatment confounding proxy and one binary outcome confounding proxy. See Section~\ref{sec_tension} for discussion of the plausibility of completeness and other base assumptions in a toy example involving a comparative effectiveness study.  Future work is needed to further translate completeness conditions into practical guidance for applications.

The necessity of $U$-relevance is intuitive: proxy variables should capture information about $U$ in order to have any hope of reducing the bias due to inability to control for $U$ in standard methods. Only variables that are reasonably believed (based on domain knowledge) to be related to $U$ (within levels of $X$ or $A$) should be considered as candidate proxy variables. In the next section, we explore how observed data can provide indirect evidence for proxy relevance (though it generally cannot be tested). 

\subsection{Empirical Proxy Selection}\label{sec_empirical}

Once candidate sets of treatment confounding proxies and outcome confounding proxies have been proposed, empirical exploration of relationships among observed variables can be used to \textit{support} (but not confirm) the plausibility that certain proxies are $U$-relevant. 

In the causal diagram typical of PCI (Figure~\ref{fig:pci_dag_labelled}), the unobserved confounders $U$ occupy a central role. In practice, $U$ is not observed, and the PCI assumptions posit the existence of such a $U$ that accounts for many of the associations observed among treatment, outcome, and proxy variables. In particular, if there exist unobserved common causes of $A$ and $Y$, and both $W$ and $Z$ are relevant to those unobserved causes $U$, we would often expect to observe the following empirical associations in the data:
\begin{itemize}
    \item $Z$ and $W$ are each associated with $Y$ conditional on $A,X$;
    \item $Z$ and $W$ are each associated with $A$ conditional on $X$; 
    \item $Z$ and $W$ are associated with each other conditional on $A,X$.
\end{itemize}

In their applied analyses, \citet{tchetgen_tchetgen_introduction_2024} and \citet{cui_semiparametric_2024} only retain proxies that are significantly associated with both the treatment and the outcome. Tests of association between $Z$ and $W$ are also recommended in \citet{cui_semiparametric_2024}, using pairwise tests for partial correlation. Not all of these associations are equally informative about $U$-relevance, a point we return to below.

These empirical checks are not \textit{sufficient} to establish $U$-relevance; one reason being that associations could also arise from violations of conditional independence assumptions. For instance, if A.\ref{condZ} ($Z\indep Y \mid U,A,X$) fails to hold, $Z$ may be associated with $Y$ given $A,X$ even in the absence of any association between $Z$ and $U$ given $A,X$. The same caveat applies to A.\ref{condW}: violations of this assumption may induce associations between $W$ and $A$ given $X$, as well as between $W$ and $Z$ given $A,X$, even in the absence of $U$-relevance. Moreover, while the absence of these associations may reflect that proxies are weakly related or unrelated to $U$, the lack of associations could also arise from minimal or no unobserved confounding---an entirely different scenario.\footnote{In this work, we generally assume that we are in a scenario where we are willing to assume unobserved confounding is a problem; else we would not be considering the use of PCI.} 

Not all of the tests of association between observed variables listed above would provide the same type of evidence for $U$-relevance. For example, even if A.\ref{condZ} and A.\ref{condW} hold, $Z$ and $A$ may be associated given $X$ even if $Z$ is not $U$-relevant, if $Z$ directly affects $A$. Similarly, an association between $W$ and $Y$ given $A,X$ may be observed even when $W$ is not $U$-relevant.
By contrast, tests of association between $Z$ and $Y$ given $A,X$, between $W$ and $A$ given $X$, and between $Z$ and $W$ given $A,X$ provide clearer evidence for or against $U$-relevance, though they require that conditional independence assumptions A.\ref{condZ} and A.\ref{condW} hold.
This is because if A.\ref{condZ} and A.\ref{condW} hold, these associations would arise through shared dependence on $U$. Nonetheless, the absence of associations between $Z$ and $A$ given $X$, or between $W$ and $Y$ given $A,X$ may still raise practical concerns about proxy relevance, and thus this full set of tests are commonly recommended as part of empirical proxy evaluation \citep{tchetgen_tchetgen_introduction_2024,cui_semiparametric_2024}.

More formal algorithms for proxy selection are presented in \citet{kummerfeld_data-driven_2022} and \citet{xie_automating_2024}. \citet{kummerfeld_data-driven_2022} proposed an algorithm for identifying \textit{disconnected} proxies (variables valid as either $W$ or $Z$) and allocating them as treatment confounding proxies or outcome confounding proxies by strength of association with treatment or outcome. \citet{xie_automating_2024} extends this beyond disconnected proxies to selection of treatment confounding proxies and outcome confounding proxies, based on either covariance structures or non-Gaussian latent factor models. Both papers assume the data distribution follows linear structural equation models, which allow proxy validity to be testable via rank conditions \citep{xie_automating_2024}. However, \citet{xie_automating_2024} does not address how to incorporate observed covariates into their algorithm and subsequent analysis, a needed extension for this method to be used in many practical settings. 

Several open questions regarding the use of such empirical tests for proxy selection and allocation should be noted. First, in practice, these procedures are often carried out using the same dataset that is used for estimation and the consequences of this practice are not fully understood. Second, empirical checks can produce false negatives: proxies may be discarded if their observed associations are weak or if the sample size is limited, and again, the implications of this are not well studied. Additionally, we emphasize these empirical tests should not replace critically reasoning about PCI assumptions using domain knowledge.

\begin{remark}\label{remark_weak_proxies}
In principle, even proxies that are only weakly associated with $U$ given $A,X$ (meaning that the measure of association between them is small in magnitude) can satisfy the completeness conditions required for identification.  This issue is reminiscent of instrumental variable methods, where identification requires only that the instrument and treatment be related, but practical guidance has emphasized the importance of ``strong'' instruments to ensure stable estimation and avoid weak-IV bias \citep{bound_problems_1995}. However, instrument strength can be tested directly, as that both the instrument and treatment as observed. Developing methods to assess and address weak proxies remains an important direction for future work.
\end{remark}

\subsection{Conceptualization of Unobserved Confounding and Tensions between Assumptions: Antidepressant Comparative Effectiveness Example}\label{sec_tension}
Thus far, we have generally assumed $U$ is clearly defined in a specific research context, but this may not always be true. In some cases, it may be tempting to define $U$ broadly as ``the collection of all unobserved confounders", making proxy conditional independence (A.\ref{condZ} and A.\ref{condW}) and latent unconfoundedness (A.\ref{latent}) feel more defensible. However, in doing so, completeness may be less likely to hold unless a sufficiently large and variable set of proxies of each type is available --- a tension also noted in \citet{cobzaru_bias_2022}. 

\begin{figure}
    \centering
    \includegraphics[width=0.5\linewidth]{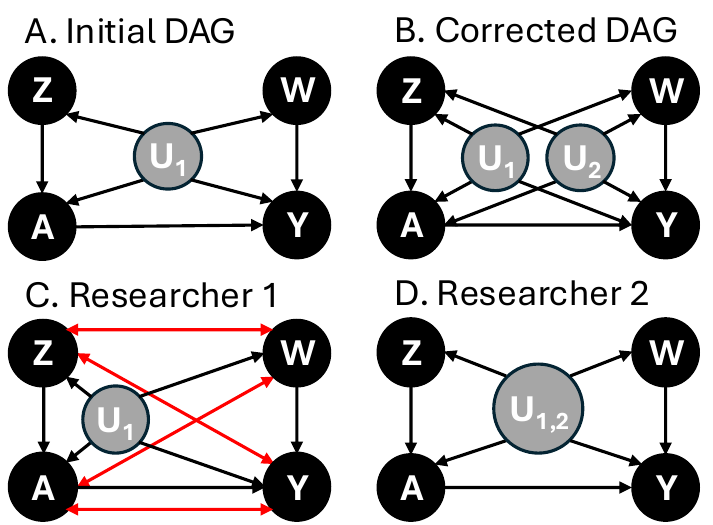}
    \caption{
Illustrative example showing how alternative representations of the unobserved confounder $U$ affect the plausibility of proximal causal inference (PCI) assumptions. 
(A) Initial directed acyclic graph (DAG) proposed by researchers. 
(B) Corrected DAG including an additional unmeasured variable $U_2$. 
(C) Researcher 1’s reformulation, treating $U_1$ as the sole unobserved confounder; replacing $U_2$ with bidirectional arrows to indicate the additional unobserved source of dependence between variables.
(D) Researcher 2’s reformulation, defining $U_{1,2} = (U_1, U_2)$ as a joint construct. }
    \label{fig:tension}
\end{figure}

As an illustrative example, suppose a team consisting of Researchers 1 and 2 posit a set of causal relationships (Figure~\ref{fig:tension}A), where $U_1$ is a single binary unobserved confounder, and $Z$ and $W$ are each a binary variable serving as a treatment confounding proxy and an outcome confounding proxy, respectively. The researchers estimate the ATE using a PCI method under this setup, with one treatment confounding proxy and one outcome confounding proxy. For this example, we omit discussion of other covariates but note that in practice there would typically be a set of covariates $X$ incorporated into estimation. Later, a collaborator suggests that a more realistic set of relationships would resemble Figure~\ref{fig:tension}B, which includes an additional binary unobserved variable $U_2$ that is a common cause of $A$, $W$, $Z$, $Y$. 

To ground this example, imagine the goal is to compare the reduction in depressive symptoms ($Y$) after taking one of two antidepressants ($A=0,1$), using observational EHR data from a cohort of patients with depression. A concern is that patients who have previously tried an antidepressant and failed to respond may be more likely to be prescribed treatment $A=1$ \citep{mattingly_effectiveness_2023} and may also be less likely to experience reduction in symptoms from the current treatment \citep{rush_acute_2006}. The researchers thus consider ``treatment resistance" as an unobserved confounder, $U_1$. For simplicity $U_1$ is considered to be a time-invariant binary variable. While medication history may be recorded in the EHR system, EHR data may be imperfect---for example, patients may receive care in other health systems \citep{lin_out--system_2018}. 
The researchers believe it is reasonable to assume the recorded history of antidepressant use prior to the index encounter where the new antidepressant was prescribed  ($Z=0,1$) only influences future depressive symptoms ($Y$) through unobserved treatment resistance ($U$), and thus use $Z$ as a treatment confounding proxy. As all participants in their cohort have either a diagnosis of recurrent major depressive disorder (MDD, denoted $W=1$) or of a single episode of MDD ($W=0$), and recurrent MDD and treatment resistance may be linked \citep{gronemann_incidence_2018}, the researchers use diagnosis of the recurrent MDD ($W=1$) relative to a single episode of MDD ($W=0$) as an outcome confounding proxy. However, a collaborator with substantive expertise points out that underlying depression is likely to be more complex than a simple binary construct and that treatment resistance and depression recurrence are related but distinct unmeasured confounders --- call them $U_1$ and $U_2$, respectively. While $W$ is a measurement of depression recurrence, it may be viewed as an imperfect given general presence of measurement error in ICD-10 codes and that a patient may not yet have had a recurrent episode. 

Which PCI assumptions are violated, assuming the updated DAG that includes $U_2$ (Figure~\ref{fig:tension}B) is accurate? Researchers 1 and 2 take different perspectives. Researcher 1 continues to think of $U_1$ as the unobserved confounder and rewrites Figure~\ref{fig:tension}B was Figure~\ref{fig:tension}C, replacing $U_2$ with bidirectional arrows. As $U_1$, $W$, and $Z$ are all binary, $W$ and $Z$ each have enough dimensions for completeness to be plausible (a necessary, though not sufficient, requirement). However, with $U=U_1$, Researcher 1 concludes that latent unconfoundedness (A.\ref{latent}: $Y(a)\indep A | U_1,X$), and proxy conditional independence assumptions (A.\ref{condZ}: $Z\indep Y \mid A,U_1,X$;  A.\ref{condW}: $W\indep (A,Z) \mid U_1,X$) are unlikely to hold since $W,A,Z,Y$ are all linked through $U_2$. 

In contrast, Researcher 2 broadens their definition of $U$ to include the joint values of $U_1$ and $U_2$. They reformulate their DAG (Figure~\ref{fig:tension}D). Under this broader definition of $U$, A.\ref{latent}, A.\ref{condZ}, and A.\ref{condW} are believed to hold. Yet completeness becomes problematic: $U$ now has four possible categories (the combinations of binary $U_1$ and $U_2$), but $W$ and $Z$ are binary and therefore cannot span all relevant variation in $U$.

Both researchers correctly identified that some PCI assumptions would be violated and thus that their estimate (obtained assuming the DAG in Figure~\ref{fig:tension}A) may be biased if the collaborator's proposed DAG (Figure~\ref{fig:tension}B) is a more accurate depiction of the true underlying relationships between observed variables. However, the way $U$ is defined---whether as a narrow or broad construct---can change which assumptions seem plausible. A lack of clarity about how $U$ is defined could lead to overlooking potentially violated assumptions. Additionally, violations of proxy conditional independence assumptions are often viewed as more serious than completeness violations --- yet, as this example shows, these assumptions are connected. 

Practically speaking, the perspective of Researcher 2 may be a more advisable position, as it suggests a path forward: adding additional proxy variables may make completeness more plausible. Still, in many settings, it may be difficult to identify even one plausible treatment confounding proxy and one outcome confounding proxy. In the antidepressant example, identifying outcome confounding proxies is particularly challenging: most pre-treatment variables recorded in the EHR data that are reasonably related to treatment resistance or recurrent depression ($U_1,U_2$) could plausibly be used directly in treatment assignment. Most post-treatment variables related to $U_1,U_2$ could plausibly be affected by treatment assignment. While this is only one use case, it shows the importance of defining the unobserved confounding mechanism and addressing the plausibility of the full set of assumptions.  

\section{Discussion}\label{sec_discussion}

PCI provides a valuable alternative to standard approaches in settings where unconfoundedness assumptions are deemed implausible. However, the validity of PCI methods generally hinges on the existence of both treatment and outcome confounding proxies that satisfy both conditional independence restrictions and completeness conditions. These assumptions are largely untestable and depend heavily on subject-matter knowledge. This parallels familiar challenges in instrumental variable and negative control analyses, where empirical associations alone cannot establish validity. We emphasize that clear reasoning about PCI assumptions requires: 
\begin{enumerate}
    \item Defining the unobserved confounder(s),
    \item Carefully thinking through the pathways linking potential proxies to the treatment and outcome, and 
    \item Considering the relative complexity of unobserved confounders and proxies (i.e., avoiding scenarios where the unobserved confounder(s) are highly complex, multidimensional processes but both sets of proxies are extremely simplified).
\end{enumerate}
When some PCI assumptions may not reasonably hold, extensions of PCI may still be available \citep{rakshit_adaptive_2025,yu_fortified_2025,ghassami_causal_2023,zhang_proximal_2023}. 

Our discussion of estimation highlights trade-offs, opportunities, and challenges. Regression-based PCI \citep{liu_regression-based_2024} and other approaches utilizing parametric forms of bridge functions \citep{cui_semiparametric_2024,miao_confounding_2024} may provide a more accessible entry point than nonparametric approaches, but depend on strong assumptions about the forms of bridge functions and proxy distributions. Additionally, closed-form bridge functions required for these methods require distributional assumptions that could be unrealistic for some settings. Nonparametric estimation offers more flexibility, but at the cost of larger data sets and computational complexity. No single estimation approach will suit all applications; instead, researchers must weigh feasibility against robustness in light of available data and plausible assumptions. Development of user-friendly software and 
implementations will be an important factor in increasing the uptake of PCI methods in practice.

As discussed throughout this work, unconfoundedness may fail if important confounders are unmeasured or measured with error. Yet nearly any covariate can be seen as an imperfect measure of some underlying construct, so the practical question may be whether the residual bias is large enough to alter conclusions. Sensitivity analyses can help gauge how strong an unmeasured confounder would need to be to change results \citep{liu_introduction_2013}. However, traditional sensitivity analysis formulas relying on simplifying assumptions (such as the bias formulas of \citet{vanderweele_bias_2011}) require careful adaptation to the PCI setting--- similar to the adaptations to differential measurement error described in \citet{rudolph_using_2018}. The recent work of \citet{huang_relative_2025} begins to address this area.

A number of related approaches offer alternatives to PCI for similar settings when unconfoundedness may not hold, though they may require additional data and often require more parametric assumptions. These alternatives include Bayesian methods \citep{hong_bayesian_2017}, latent variable modeling \citep{nguyen_propensity_2020}, and calibration strategies \citep{sturmer_adjusting_2005}. Comparative work examining the trade-offs between these approaches, various PCI estimators, and standard estimators assuming unconfoundedness could aid practitioners in determining which methods are most appropriate in particular settings. 

In conclusion, PCI is a promising framework to address unobserved confounding through use of proxy variables. By emphasizing confounding mechanisms, clarifying trade-offs in estimation, and discussing how assumptions map onto a specific example, this work aims to support thoughtful and transparent use of PCI as part of the broader causal inference toolkit.

\begin{appendix}
\renewcommand{\thesection}{\Roman{section}} 
\renewcommand{\thesubsection}{\Roman{section}.\Roman{subsection}} 

\section{PCI Assumptions and Identification}\label{appendix_pci}

\subsection{Proxies in the Language of Measurement Error}\label{appendix_measurement}

\begin{figure}
    \centering
    \includegraphics[width=.25\linewidth]{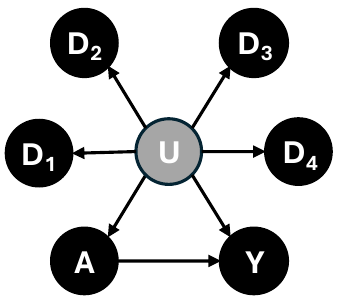}
    \caption{A DAG that depicts an example of a traditional latent variable model, with four measurements. These can also be viewed as \textit{disconnected} proxies.}
    \label{fig:disconnected}
\end{figure}

To further examine the connection between ``measurements" and proxies, consider a latent variable model in which we have access to $J$ variables that can all be considered imperfect measurements of $U$, denoted $\{D_1,D_2,...D_J\}$. The DAG in Figure~\ref{fig:disconnected} (with $J=4$) depicts a special case: that all measurement error is \textit{non-differential}, meaning: $D_j \perp Y \mid A, U$ and $D_j \perp A \mid U$ for all $j=1,...,J$ and \textit{independent }$D_j \perp D_k\mid A, U$ for all $j\not=k$ \cite{vanderweele_results_2012}. In this case, these measurements can be viewed as \textit{disconnected proxies} (see Remark~\ref{remark_disconnected}), as they are each reasonable candidates for use as either a treatment confounding proxy or an outcome confounding proxy. That is, based on this DAG, assumptions A.\ref{condZ} and A.\ref{condW} would be satisfied by any partitioning of measurements into two sets, so long as we have at least one $Z$ and at least one $W$. For example, we may choose $Z=\{D_1,D_2\}$, $W=\{D_3,D_4\}$. However, subsequent assumptions need to be considered in these allocations: in particular, the completeness conditions (which differ slightly based on identification approach) on whether both sets of $Z$ and $W$ have \textit{sufficient variability} relative to $U$ (and to each other). Completeness may be more reasonable under some partitions. For example, it may be more reasonable to assume that $Z$ is complete for $U$ with $Z=\{D_1,D_2\}$ than with $Z=\{D_1\}$. Methods for allocating disconnected proxies to be used as a particular type of proxy are discussed further in Section~\ref{sec_empirical}. 

General PCI scenarios and latent variable models are less restrictive (i.e., allow more dependence) than the non-differential measurement error case described above. Still, the PCI conditional independence assumptions A.\ref{condZ} and A.\ref{condW} admit a natural interpretation in the language of measurement error:
\begin{enumerate}
    \item A valid treatment confounding proxy (satisfying $Z \indep Y \mid U,A, X$) could also be called a measurement of $U$ with non-differential error \textit{with respect to $Y$}, conditional on $A,X$
    \item A valid outcome confounding proxy (satisfying $W \indep A \mid U, X$) could be called a measurement of $U$ with non-differential measurement error \textit{with respect to $A$}, conditional on $X$.
\end{enumerate}

Whether such language would be used is largely context dependent. Typically, the work ``measurement" evokes the idea that $Z$ and $W$ are attempting to quantify $U$, with some error. ``Proxy" is a more general term that also encapsulates negative controls. See section~\ref{sec_practical} for further discussion. 

\subsection{PCI Assumptions in a  Parametric Model}\label{ex:combo}

Here we provide an example of how the base PCI assumptions map onto the parameters of a fully parametric data generating model (DGM). Similar to the DGM used in the simulations of \citet{cui_semiparametric_2024}, consider data generated from the following model, where all random variables other than $U$ are observed.

\[\begin{array}{l}
X \sim N(0,1),\\
A \mid X \sim \text{Bernoulli}\left(p_a = \text{expit}(\alpha_0 + \alpha_x X)\right)\\
U \mid X, A \sim N(\mu_0 + \mu_a A + \mu_x X, \sigma_{u|xa}^2),\\
(Z, W) \mid U, A, X \sim\\ \quad \quad MVN \left(
\begin{pmatrix}
\theta_0 + \theta_a A + \theta_u U + \theta_x X \\
\omega_0 + \omega_a A + \omega_u U + \omega_x X
\end{pmatrix},
\Sigma
\right),\\
Y \mid A,X,U,Z,W \sim N(\mu_{y|axuzw},\sigma^2_{y|axuzw}),\\

\end{array}
\]
where:
\[
\begin{array}{l}
\Sigma=\begin{pmatrix}
\sigma_{z|xau}^2 & \sigma_{zw|xau} \\
\sigma_{zw|xau} & \sigma_{w|xau}^2 \end{pmatrix},\\
\mu_{y|axuzw}=\beta_0 + \beta_aA + \beta_uU + \beta_xX + \beta_zZ + \beta_wW
\end{array}
\]

\textbf{Consistency} (A.\ref{cons}) holds by construction as $Y$ is generated as a function of $A, U, Z, W$, and an error term that is independent of these variables.

\textbf{Latent Positivity} holds as long as $\alpha_0,\alpha_x,\mu_0,\mu_a,\mu_x$ are finite and $\sigma_{u|xa}^2>0$. 

\textbf{Conditional independence assumptions} $W \indep (Z,A) \mid U,X$ (A.\ref{condW}) and  $Z \indep Y \mid U,X,A$ (A.\ref{condZ}) hold if $\sigma_{wz|uax}=0$, $\omega_a=0$, and $\beta_z=0$.

\textbf{Unconfoundedness does not hold.} While this is not a formal assumption of PCI, it would be unnecessary to use a method like PCI if unconfoundedness (A.\ref{exc}) does hold--- thus, it is worth briefly justifying. Although $U$ is modeled statistically as a function of $A, X$ in the DGM, the causal interpretation for proximal causal inference treats $U$ as a latent confounder---a common cause of both $A$ and $Y$. In this DGM, this maps onto the parameter restrictions $\mu_a \not= 0$ and $\beta_u \not= 0$, which imply that $U$ influences both treatment assignment and the outcome. Since $U$ is unobserved, the standard unconfoundedness assumption $Y(a) \indep A \mid X$ does not hold.

\textbf{Latent Unconfoundedness} holds (assuming the conditional independence assumptions A.\ref{condZ} and A.\ref{condW}) hold because under the causal interpretation of $U$ as a latent confounder, the only common causes of $A$ and $Y$ are $U,X$.

\textbf{Completeness Conditions} generally require sufficient variability in $Z$ and/or $W$ relative to $U$ or to each other.
First note that given A.\ref{condZ} and A.\ref{condW}, $W$ is $U$-relevant (not independent of $U$ conditional on $A,X$) in this DGM if $\omega_u \neq 0$. Similarly, $Z$ is $U$-relevant if $\theta_u \neq 0$.
Second, note that $Z$, $W$, and $U$ are all univariate in this DGM---thus have the same number of dimensions. To formally justify completeness conditions (A.\ref{compZW}, A.\ref{compUZ}, A.\ref{compWZ}, and/or A.\ref{compUW}) one can invoke results for completeness of exponential families \citep{newey_instrumental_2003}. See \citet{miao_identifying_2023} for further explanation, examples, and citations to the broader completeness literature.

\subsection{Example of a Completeness Failure}\label{ex_comp_fail}
To illustrate a case where completeness of $Z$ for $U$ fails due to insufficient dimensionality (despite both being normally distributed) suppose that:
$$U = (U_1, U_2)^T \mid A,X\sim N\!\left((0,0)^T, I_2\right),$$
$$Z \mid U,A,X \sim N\left(\theta_1 U_1 + \theta_2 U_2,\, \sigma^2_{z\mid uax}\right)$$
where $\theta_1,\theta_2 \neq 0$ and $\sigma^2_{z\mid uax} > 0$. 
Thus, $Z$ is associated with both components of $U$ conditional on $A,X$ but is one-dimensional, while $U$ is two-dimensional.

One way to show that $Z$ is not complete for $U$ (A.\ref{compUZ}) in this case is to find one function $g(U)$ where $\mathbb{E}[g(U) \mid Z,A,X]=0 $ does not imply  $g(U)=0$.

We select $g(U) = \theta_2 U_1 - \theta_1 U_2$.

First, note that as the joint distribution of $(U,Z)\mid A,X$ is multivariate normal, the conditional expectation is:
\begin{align*}
    \mathbb{E}[U \mid Z,A,X] &= 
\mathrm{Cov}(U,Z\mid A,X)\,\mathrm{Var}(Z\mid A,X)^{-1}Z\\
&=(\theta_1,\theta_2)^T
(\theta_1^2 + \theta_2^2+\sigma^2_{z\mid uax})^{-1}Z
\end{align*}

Therefore,
\begin{align*}
    \mathbb{E}[g(U) \mid Z,A,X] &=\mathbb{E}[(\theta_2U_1 - \theta_1U_2) \mid Z,A,X] \\
&= (\theta_1^2 + \theta_2^2+\sigma^2_{z\mid uax})^{-1}\,(\theta_2\theta_1 - \theta_1\theta_2)\,Z \\&=0.
\end{align*}

However, $g(U)=\theta_2 U_1 - \theta_1 U_2$ is not almost surely zero since
$\mathrm{Var}(g(U)\mid A,X) = \theta_1^2 + \theta_2^2 > 0$. This demonstrates how completeness can fail when proxy dimensionality is insufficient ($\dim(Z)<\dim(U)$).

\subsection{Walk-through of Identification with Outcome Bridge Function}\label{appendix_h_ident}

Here we summarized the identification results first shown in \citet{miao_identifying_2018}, in the potential outcome framework as described in \citet{tchetgen_tchetgen_introduction_2024}. The first step is to show that the base PCI assumptions, plus the assumption of the existence of an outcome bridge function (A.\ref{PCIh}) and that $Z$ is complete for $U$ (A.\ref{compUZ}) imply Eq. \ref{PCIhu}:
\begin{equation*}
 \mathbb{E}[Y\mid A,X,U] = \int_w h(w,A,X)f(w|A,X,U)du
\end{equation*}

To show this, begin with the definition of the outcome bridge, as assumed to exist in A.\ref{PCIh}, satisfying Eq. \ref{PCIh:eq}:
\begin{equation*}
 \mathbb{E}[Y\mid Z,A,X] = \int_w h(w,A,X)f(w|Z,A,X)dw
\end{equation*}

Starting with the left hand side, taking the iterated expectation, and using A.\ref{condZ} ($Z\indep Y\mid U,A,X$), we have:
\begin{align*}
    \mathbb{E}[Y \mid A,X,Z] &= \mathbb{E}[\mathbb{E}[Y\mid A,X,Z,U]|A,X,Z]\\
    &= \int_u \mathbb{E}[Y\mid A,X,Z,u]f(u|A,X,Z)du \\
    &= \int_u\mathbb{E}[Y\mid A,X,u]f(u|A,X,Z)du 
\end{align*}

Then, using the right hand side, one can similarly take the iterated expectation and use A.\ref{condW}:
\begin{align*}
    &\int_w h(w,A,X)f(w|Z,A,X)dw =\\ 
    &= \int_u \int_w h(w,A,X)f(w|Z,A,X,u)f(u|Z,A,X)dwdu\\
    &= \int_u \int_w h(w,A,X)f(w|A,X,u)f(u|Z,A,X)dwdu
\end{align*}

Setting both sides equal, we have:
\begin{align*}
\int_u \mathbb{E}&(Y\mid A,X,u)f(u|Z,A,X)du =\\&= \int_u \bigl[\int_wh(w,A,X)f(w|A,X,u)dw\bigr]f(u|Z,A,X)du
\end{align*}

By the completeness condition  (A.\ref{compUZ}), this implies Eq.~\ref{PCIhu}. To see this directly, put both terms on the same side of the equation and let 
$$g(u) = \mathbb{E}[Y\mid A,X,u] - \int_wh(w,A,X)f(w|A,X,u)dw.$$
Then we have $\int_u g(u)f(u|Z,A,X)du=0$ which by completeness (A.\ref{compUZ}) implies $g(u)=0$, showing the desired equality, Eq. \ref{PCIhu}.

Now, Eq. \ref{PCIhu}, along with the other core PCI assumptions, can be used to show the identification of $E(Y(a))$. Start with $\mathbb{E}[Y(a)]$, take an iterated expectation twice, apply A.\ref{latent}, A.\ref{cons}, and Eq. \ref{PCIhu}, to obtain:
\begin{align*}
    \mathbb{E}&[Y(a)] = \int_{x}\int_{u}  \mathbb{E}[Y(a) \mid u,x] f(x,u)dudx \\
    &= \int_{x}\int_{u}  \mathbb{E}[Y(a) \mid x,u,a] f(x,u)dudx \\
    &= \int_{x}\int_{u}   \mathbb{E}[Y \mid x, u, a] f(x,u)dudx \\
    &= \int_{x}\int_{u}\bigl[\int_{w} h(w,a,x)f(w|x,u)dw\bigr]f(x,u)dudx\\
    &= \int_{x}\int_w h(w,a,x)f(w|x)f(x)dwdx\\
    &=\mathbb{E}[h(W,a,X)]
\end{align*}

Note that the assumption of latent positivity (A.\ref{latpos}) is what allows  $\mathbb{E}[Y \mid x, u, a]$ to be defined for all combinations of $x, u, a$ that have positive probability.

\section{Estimation of the ATE}\label{appendix_estimation}
\subsection{Further Detail on Estimating Equations for a Parametric Outcome Bridge Function}\label{appendix_m_est}

The framework of M-estimation \citep{stefanski_calculus_2002} can be used to explain the general structure of estimation in PCI, when assuming a parametric form of the outcome bridge function. Let $D_i$ represent observed i.i.d. data. In M-estimation, the parameters of interest $\theta$ solve an \textit{unconditional moment condition} of the form $\mathbb{E}[f(D_i,\theta_0)]=0$ (where $\theta_0$ is the truth) and the corresponding M-estimator $\hat\theta$ is defined by $P_n[f(D_i,\hat\theta)]=0$ in the data \citep{stefanski_calculus_2002}.\footnote{There are other more general definitions of ``M-estimation" as an estimator solving a minimizing or maximizing problem---we use the definition of \citet{stefanski_calculus_2002} here.}

The definition of the outcome bridge function (Equation~\ref{PCIh:eq} in A.\ref{PCIh}) is a \textit{conditional} moment equation. A first step is to convert this into unconditional moments suitable for M-estimation. First note that Equation~\ref{PCIh:eq} can be rearranged as:
$$\mathbb{E}[Y - h(W,A,X;\eta)\mid Z,A,X]=0.$$

Then, using properties of conditional expectations, this implies:\footnote{A quick justification that $E[U|V]=0$ implies $E[k(V) U]=0$: $E[k(V) U]=E[E[k(V) U|V]]=E[k(V)E[ U|V]]=0$}
 \begin{equation}\label{eq:moment_h}
     \mathbb{E}[(Y - h(W,A,X;\eta))k(Z,A,X)]=0.
 \end{equation}
 The vector $k(Z,A,X)$ is chosen by the user but must have dimension greater than or equal to the dimension of $\eta$. With the simple linear $h(W,A,X,\eta)$ (Eq. \ref{eq:linear_h}), a natural choice is $k(Z,A,X)=(1,Z,A,X)^T$. This is the choice presented in the main text. 

 Thus, we can define the following estimating functions, as in \citet{miao_confounding_2024}:
 \begin{equation}\label{eq:h_gmm}
     f_h(D_i,\theta) = \begin{bmatrix}
      (Y_i - h(W_i,A_i,X_i;\eta_h))k(Z_i,A_i,X_i)\\
      \tau - (h(W_i,1,X_i;\eta_h)-h(W_i,0,X_i;\eta_h))
  \end{bmatrix}
 \end{equation}

where $\theta=(\eta,\tau)$ and $D_i$ represents the observed data, $D_i=(X_i,Z_i,W_i,A_i,Y_i)$, which are assumed to be i.i.d..
 The last line simply specifies the ``proximal g-computation" step, coming from $\tau_h=E[h(W,1,X)-h(W,0,X)]$. 
 The M-estimator for the vector of parameters $\theta$ (of which the ATE, $\tau$, is the last element), is defined as the solution to $P_n[f_h(D_i,\hat\theta)]=0.$ 

If the number of equations here is the same as the dimension of $\theta$, the system can be solved directly (see the example in Section~\ref{sec_estimation}) and variance can be estimated using standard M-estimation sandwich estimators \citep{stefanski_calculus_2002}. 

In cases with more equations than unknowns, we can use generalized method of moments (GMM) \citep{newey_chapter_1994, harris_introduction_1999} to estimate $\hat{\theta}$--- this is the approach laid out in \citet{miao_confounding_2024}. Defining $G_n(\theta)=P_n[f_h(D_i,\theta)]$, the GMM estimator\footnote{This can also be called an M-estimator.} $\hat{\theta}$ is:
$$\hat\theta=\text{argmin}_\theta G_n(\theta)^T\Omega G_n(\theta)$$
where $\Omega$ is a weighting matrix defined by the user. Under regularity conditions, the GMM estimator $\hat\theta$ is consistent and asymptotically normal and the form of the variance of $\hat\theta$ (and thus, of $\hat\tau$) is provided in \citet{miao_confounding_2024}.

Noting that $(Y_i-h(W_i,A_i,X_i))$ and $\tau - (h(W_i,1,X_i;\eta_h)-h(W_i,0,X_i;\eta_h))$ are both scalar, the number of estimating equations in Eq. \ref{eq:h_gmm} is dictated by the form of $k(Z_i,A_i,X_i)$. A natural question is: what form of $k(Z_i,A_i,X_i)$ minimizes the asymptotic variance of $\hat\theta$?  Results from \citet{newey_chapter_1994} can be applied to the specific case of Equation \ref{PCIh:eq} to show that the optimal $k$ has the form:

\begin{align*}
    k_{opt}(Z,A,X)&=\mathbb{E}[-\nabla_{\eta}\ h(W,A,X;\eta)|Z,A,X]^T\\
    &\cdot\mathbb{E}[(Y-h(W,A,X;\eta))^2|Z,A,X]^{-1}
\end{align*}

 Note that this $k_{opt}$ will have the same dimension as $\eta$ by definition.

 \citet{cui_semiparametric_2024} approach estimation and efficiency from a broader, semiparametric perspective. They derive the general form of the influence functions for all regular and asymptotically linear estimators of the bridge function parameters (for both the outcome bridge and treatment bridge functions), as well as the form of the efficient influence function (see appendix of \citep{cui_semiparametric_2024} Theorem E.1 and E.2). This leads to the same form of $k_{opt}$ as provided above, but framed as a component of the efficient influence function.
 
 Note that operationalizing the most efficient form of estimation (i.e. by finding $k_{opt}$) in a particular case would require estimating additional components of the observed data distribution, and additional assumptions about $\mathbb{P}(Y,W\mid Z,A,X)$ beyond the parametric form of $h(W,A,X)$ would be needed \citep{cui_semiparametric_2024}.\citet{cui_semiparametric_2024} notes that the efficiency gains that are made in practice by making these additional assumptions may be modest.

\subsection{General Two-Stage ATE Estimation}\label{appendix_g_comp}
The generalization of the two-stage approach of \citet{tchetgen_tchetgen_introduction_2024} requires assuming not only a parametric form $h(W,A,X;\eta)$ but making assumptions on the conditional distribution of the outcome confounding proxies $f(W\mid Z,A,X;\theta)$ \citep{tchetgen_tchetgen_introduction_2024}. The general two-stage procedure is then \citep{tchetgen_tchetgen_introduction_2024}: 
\begin{enumerate}
    \item Obtain $\hat\theta$, an estimate of $\theta$, using assumptions on $f(W\mid Z,A,X; \theta)$ (e.g., by maximum likelihood estimation). 

    \item Obtain $\hat\eta$, an estimate of $\eta$, by regressing $Y$ on $\mathbb{E}[Y \mid Z,A,X; \hat\theta,\eta]$, where $\hat\theta$ is plugged in for $\theta$ to evaluate 
\begin{align*}
    \mathbb{E}[Y \mid& Z,A,X; \hat\theta,\eta]=\int_wh(w,a,x; \eta)f(W\mid Z,A,X; \hat\theta)dw.
\end{align*}
\end{enumerate}

Then, the ATE is obtained by plugging in $\hat\eta$: $\hat\tau_h=P_n[h(W,1,X;\hat\eta)-h(W,0,X;\hat\eta)].$

When $\mathbb{E}[Y \mid Z,A,X; \theta,\eta]$ and Eq.\ref{PCIh:eq}) has a \textit{closed form} (meaning it can be simplified such that it no longer involves marginalization over $w$), this procedure may be straightforward. In the special case of a linear $h(W,A,X;\eta)$, specific assumptions on $f(W\mid Z,A,X)$ are not required for a closed form $\mathbb{E}[Y \mid Z,A,X;\eta]$. 
In more complex cases, such as binary $Y$ with the outcome bridge function assumed to have a logistic form ($\logit(h(W,A,X))=\vecWAX\eta$), particular assumptions on the distribution $f(W\mid Z,A,X; \theta)$ (involving a logistic bridge function probability density \citep{wang_matching_2003}) are required to yield a logit form for $\mathbb{E}[Y \mid Z,A,X; \theta,\eta]$--- see the appendix of 
\citet{tchetgen_tchetgen_introduction_2024} for specifics. 

If we were to make assumptions on the form of $h(W,A,X;\eta)$ and/or $f(W\mid A,X,Z;\theta)$ that do not yield a closed-form $\mathbb{E}[Y\mid Z,A,X;\eta,\theta]$, numeric methods such as Monte Carlo approximation would be required to implement this approach \citep{tchetgen_tchetgen_introduction_2024}. Implementation details and properties of the ATE estimator in such cases have not yet been well-documented. 

\citet{tchetgen_tchetgen_introduction_2024} suggests quantifying the uncertainty of these estimators using bootstrap variance estimators. Formal conditions guaranteeing the consistency of the bootstrap in these settings remain unclear. However, when the procedure yields a two stage regression that can be written as a set of estimating equations, known variance results can be used to construct standard errors. 

\subsection{The Very Special Case: Two-Stage Linear Regression}\label{appendix_very_special}

Two-stage linear regression in PCI has been described in depth in several recent works, including \citet{tchetgen_tchetgen_introduction_2024}, \citet{zivich_introducing_2023}, and \citet{liu_regression-based_2024}. There are various ways to motivate such an approach. For example, one may assume the following structural equations \citep{tchetgen_tchetgen_introduction_2024,zivich_introducing_2023,liu_regression-based_2024}:
\begin{align*}
\mathbb{E}[Y \mid A, Z, X, U] &= \gamma_0 + \gamma_a A + \gamma_u U + \gamma_x X, \\
\mathbb{E}[W \mid A, Z, X, U] &= \omega_0 + \omega_u U + \omega_x X,
\end{align*}
with $\omega_u \neq 0$. Taking iterated expectations yields:
$$
\mathbb{E}[W \mid A, Z, X] 
= \omega_0 + \omega_u \mathbb{E}[U \mid A, Z, X] + \omega_x X,
$$
which, rearranging, gives:
$$
\mathbb{E}[U \mid A, Z, X] 
= \frac{1}{\omega_u}\left\{\mathbb{E}[W \mid A, Z, X] - \omega_0 - \omega_x X\right\}.
$$
Substituting this expression into the iterated expectation of the outcome model, we have:
\begin{align*}
\mathbb{E}[Y \mid A, Z, X] 
&= \gamma_0 + \gamma_a A + \gamma_u \mathbb{E}[U \mid A, Z, X] + \gamma_x X \\
&= \eta_0 + \eta_a A + \eta_w \mathbb{E}[W \mid A, Z, X] + \eta_x X,
\end{align*}
where
\[
\eta_a=\gamma_a,\quad
\eta_0 = \gamma_0 - \frac{\gamma_u \omega_0}{\omega_u}, 
\quad
\eta_w = \frac{\gamma_u}{\omega_u}, 
\quad
\eta_x = \gamma_x - \frac{\gamma_u \omega_x}{\omega_u}.
\]

This has the exact same form for $\mathbb{E}[Y \mid A, Z, X]$ as if $h(W, A, X; \eta) = \eta_0 + \eta_a A + \eta_x X + \eta_w W$ is assumed directly. 

A further specialization commonly used, as in \citet{liu_regression-based_2024} is to also fit a linear model as for the first-stage regression, assuming:
$$
\mathbb{E}[W \mid A, Z, X; \theta] = \theta_0 + \theta_a A + \theta_x X + \theta_z Z,
$$
leading to the familiar two-stage linear regression procedure. This can be written as first solving: 
$$P_n[(W-(\theta_0 + \theta_a A + \theta_x X + \theta_z Z)(1,Z,A,X)^T]=0$$ to estimate $\hattheta$ and then solving 
$$P_n[(Y-(\eta_0 + \eta_aA + \eta_w\widehat{W} + \eta_xX))(1,\widehat{W},A,X)^T)=0$$
to estimate $\eta$, where $\widehat{W}=\hat\theta_0 + \hat\theta_a A + \hat\theta_x X + \hat\theta_z Z$ using the first stage regression. 

It may not be immediately obvious but such a procedure actually yields the exact same estimates as solving the estimating equations in Section~\ref{sec_estimation}, Eq. \ref{eq:pci_h_ee}:
$$P_n\big[(Y-h(W,A,X;\eta))(1,Z,A,X)^T]=0$$

The key observation is that the fitted values $\widehat W$ from the first-stage regression are the linear projection of $W$ onto the space spanned by $(1,Z,A,X)$. Thus $\widehat W$ is a linear combination of $(1,Z,A,X)$, and the residual $W-\widehat W$ is orthogonal to each of $1$, $Z$, $A$, and $X$.

Writing $\widehat W = W-(W-\widehat W)$ and substituting into the second-stage estimating equations yields
\begin{align*}
    P_n\big[(Y-(\eta_0+\eta_a A+&\eta_w W+\eta_x X)
+ \eta_w (W-\widehat W))\\ &\cdot(1,\widehat W,A,X)^T]=0.
\end{align*}

The term involving $W-\widehat W$ vanishes because $W-\widehat W$ is orthogonal to the space spanned by $(1,Z,A,X)$, which contains $(1,\widehat W,A,X)$. Therefore the estimating equations reduce to
$$P_n\big[(Y-(\eta_0+\eta_a A+\eta_w W+\eta_x X))(1,\widehat W,A,X)^T]=0.$$

Since $\widehat W$ itself lies in the span of $(1,Z,A,X)$, it can be written as a linear combination of $(1,Z,A,X)$, and so we can write:
$$P_n\big[(Y-(\eta_0+\eta_a A+\eta_w W+\eta_x X))\Theta\,(1,Z,A,X)^T\big]=0$$
where $(1,\widehat W,A,X)^T=\Theta(1,Z,A,X)^T$ and $\Theta$ is a constant matrix. Thus, the estimating equations arising from the two-stage procedure correspond to linear combinations of the original estimating equations $P_n\big[(Y-(\eta_0+\eta_a A+\eta_w W+\eta_x X))(1,Z,A,X)^T\big]=0.$
These equations enforce that the residual $Y-(\eta_0+\eta_a A+\eta_w W+\eta_x X)$ is orthogonal to the linear span of $(1,Z,A,X)$, and in particular to $(1,Z,A,X)^T$ itself. Therefore, either set of estimating equations yields the same estimator $\hat\eta$.

\subsection{Treatment Bridge Function Estimating Equations}\label{ee_q_derivation}

In this section we show how the definition of the treatment bridge function (Eq. \ref{PCIq:eq} in A.\ref{PCIq}) can be used to justify the estimating equations of the form in Eq. \ref{eq:pci_q_ee}. 

Beginning with the definition of the treatment bridge function: 
\begin{equation*}
\mathbb{E}[ q(Z,a,X) \mid W, A=a, X ] = \frac{1}{\mathbb{P}(A=a \mid W, X)}.
\end{equation*}

The left hand side is equal to 
\[
\mathbb{E}\!\left[ q(Z,a,X) \frac{I(A=a)}{\mathbb{P}(A=a \mid W,X)} \mid W, X \right],
\]
so this implies
\[
\mathbb{E}[ I(A=a) q(Z,a,X) \mid W,X ] = 1.
\]
Stacking this expression for both $a=0$ and $a=1$, we have:
\begin{equation}\label{q_step_3}
\mathbb{E}\!\left[
\begin{pmatrix}
A q(Z,A,X)\\
(1-A) q(Z,A,X)
\end{pmatrix}
\Bigg| W,X
\right]
=
\begin{pmatrix}
1\\
1
\end{pmatrix},
\end{equation}

which is equivalent to

\begin{equation}\label{q_step_4}
\mathbb{E}\!\left[
\begin{pmatrix}
A q(Z,A,X) - 1\\
A q(Z,A,X) - (1-A) q(Z,A,X)
\end{pmatrix}
\Bigg| W,X
\right]
=
\begin{pmatrix}
0\\
0
\end{pmatrix},
\end{equation}

where the new second element is the difference between the two elements in Eq. \ref{q_step_3}. 

Eq. \ref{q_step_4} implies:

\begin{equation}\label{q_step_5}
\mathbb{E}\left[
\begin{pmatrix}
A q(Z,A,X) - 1\\
A q(Z,A,X) - (1-A) q(Z,A,X)
\end{pmatrix}
k(W,X)
\right]
=
\begin{pmatrix}
0\\
0
\end{pmatrix},
\end{equation}

for arbitrary function \( k(W,X) \).

Note that
\[
(-1)^{1-A} q(Z,A,X)
= A q(Z,A,X) - (1-A) q(Z,A,X),
\]
\[
(-1)^{1-A} q(Z,A,X)\, A = A q(Z,A,Z),
\]

so Eq. \ref{q_step_5} can be written as:
\[
\mathbb{E}\!\left[
(-1)^{1-A} q(Z,A,X)
\begin{pmatrix}
A\\[6pt]
k(W,X)
\end{pmatrix}
-
\begin{pmatrix}
1\\
0
\end{pmatrix}
\right]
=
\begin{pmatrix}
0\\
0
\end{pmatrix}.
\]

Now let \( k(W,X) = (1, W, X)^{T} \) and we have

\[
\mathbb{E}\!\left[
(-1)^{1-A} q(Z,A,X)
\begin{pmatrix}
1\\
W\\
A\\
X
\end{pmatrix}
-
\begin{pmatrix}
0\\
0\\
1\\
0
\end{pmatrix}
\right]
=
\begin{pmatrix}
0\\
0\\
0\\
0
\end{pmatrix},
\]

which justifies the estimation equations (Eq. \ref{eq:pci_q_ee}):

\[
P_n\left[
(-1)^{1-A} q(Z,A,X)
\begin{pmatrix}
1\\
W\\
A\\
X
\end{pmatrix}
-
\begin{pmatrix}
0\\
0\\
1\\
0
\end{pmatrix}
\right]
=
\begin{pmatrix}
0\\
0\\
0\\
0
\end{pmatrix},
\]

as used in \citet{cui_semiparametric_2024}.

\subsection{Binary Nonparametric Estimation Example}\label{ex_binary_np}
When all variables are binary or categorical, nonparametric estimation is equivalent to fitting a saturated model for the bridge functions \citep{shi_multiply_2020}. Consider the case when $A,W,Z,U,X$ are all binary variables and $h(W,A,X)$ is essentially a look-up table with $2^3=8$ values in it. For a particular $A=a,X=x$, we can write Eq. \ref{PCIh:eq} as:
$$y_{ax} = \psi_{ax} h_{ax}$$

where 
$y_{ax} = [E(Y|Z=0,a,x), E(Y|Z=1,a,x)]^T$, 
$$\psi_{ax}=\begin{bmatrix}
P(W=0|Z=0,x,a) & P(W=1|Z=0,x,a) \\
P(W=0|Z=1,x,a) & P(W=1|Z=1,x,a)
\end{bmatrix},
$$ and $h_{ax}=[h(W=0,a,x), h(W=1,a,x)]^T$ 

If $\psi_{ax}$ is invertible, we can solve for $h_{ax}$:
$$\psi_{ax}^{-1}y_{ax} = h_{ax}.$$

The condition that $\psi_{ax}$ is invertible is equivalent to the completeness conditions A.\ref{compUZ} and A.\ref{compZW} for binary or categorical data, as noted in \citet{miao_identifying_2018}.

We can estimate $y_{ax}$ and $\psi_{ax}^{-1}$ directly using the data. For example, $\hat{\mathbb{P}}(W=1\mid Z=0,X=0,A=1)$ is simply the fraction of data points that have $W=1$ among those with $Z=0,X=0,A=1$. Then, the ATE is estimated as: 
$\hat{\tau}_{h} = \mathbb{P}_n[\hat{h}(W,a=1,X)  - \hat{h}(W,a=0,X)]$

The same procedure can be done with the treatment confounding bridge function.

\section{Acknowledgments}
The authors would like to thank Yifan Cui and the authors of \textit{Semiparametric Proximal Causal Inference} for sharing their simulation code; to Peter Liu, Helen Guo, Jessie Tong, Betsy Ogburn, and Eric Kernfeld for helpful discussion.

Authors' additional affiliations: Trang Q. Nguyen has a joint appointment in the Department of Biostatistics, Johns Hopkins Bloomberg School of Public Health (BSPH), Maryland, USA. Peter P. Zandi has joint appointments in the Department of Epidemiology and Department of Mental Health, BSPH. Elizabeth A. Stuart has joint appointments in the Department of Health Policy and Management and Department of Mental Health, BSPH. Harsh Parikh is an Affiliate of the Department of Biostatistics, BSPH. 

\subsection{Funding}
Grace V. Ringlein was supported by NIMH R01MH126856 (PI: Stuart). 

\end{appendix}

\bibliography{references}

@article{miao_identifying_2018,
    title = {Identifying causal effects with proxy variables of an unmeasured confounder},
    volume = {105},
    copyright = {https://academic.oup.com/journals/pages/open\_access/funder\_policies/chorus/standard\_publication\_model},
    issn = {0006-3444, 1464-3510},
    url = {https://academic.oup.com/biomet/article/105/4/987/5073056},
    doi = {10.1093/biomet/asy038},
    language = {en},
    number = {4},
    urldate = {2024-08-06},
    journal = {Biometrika},
    author = {Miao, Wang and Geng, Zhi and Tchetgen Tchetgen, Eric J},
    month = {dec},
    year = {2018},
    pages = {987--993},
}

@article{kuroki_measurement_2014,
    title = {Measurement bias and effect restoration in causal inference},
    volume = {101},
    issn = {0006-3444, 1464-3510},
    url = {https://academic.oup.com/biomet/article-lookup/doi/10.1093/biomet/ast066},
    doi = {10.1093/biomet/ast066},
    language = {en},
    number = {2},
    urldate = {2024-09-12},
    journal = {Biometrika},
    author = {Kuroki, M. and Pearl, J.},
    month = jun,
    year = {2014},
    pages = {423--437},
}

@article{lipsitch_negative_2010,
    title = {Negative {Controls}: {A} {Tool} for {Detecting} {Confounding} and {Bias} in {Observational} {Studies}},
    volume = {21},
    issn = {1044-3983},
    shorttitle = {Negative {Controls}},
    url = {https://journals.lww.com/00001648-201005000-00017},
    doi = {10.1097/EDE.0b013e3181d61eeb},
    language = {en},
    number = {3},
    urldate = {2024-08-06},
    journal = {Epidemiology},
    author = {Lipsitch, Marc and Tchetgen Tchetgen, Eric and Cohen, Ted},
    month = may,
    year = {2010},
    pages = {383--388},
}

@article{tchetgen_tchetgen_introduction_2024,
    title = {An {Introduction} to {Proximal} {Causal} {Inference}},
    volume = {39},
    issn = {0883-4237},
    url = {https://projecteuclid.org/journals/statistical-science/volume-39/issue-3/An-Introduction-to-Proximal-Causal-Inference/10.1214/23-STS911.full},
    doi = {10.1214/23-STS911},
    number = {3},
    urldate = {2024-07-18},
    journal = {Statistical Science},
    author = {Tchetgen Tchetgen, Eric J. and Ying, Andrew and Cui, Yifan and Shi, Xu and Miao, Wang},
    month = aug,
    year = {2024},
}

@misc{ghassami_causal_2023,
    title = {Causal {Inference} with {Hidden} {Mediators}},
    url = {http://arxiv.org/abs/2111.02927},
    doi = {10.48550/arXiv.2111.02927},
    urldate = {2025-05-19},
    publisher = {arXiv},
    author = {Ghassami, AmirEmad and Yang, Alan and Shpitser, Ilya and Tchetgen Tchetgen, Eric},
    month = jan,
    year = {2023},
    note = {arXiv:2111.02927 [math]},
}

@misc{li_doubly_2022,
    title = {Doubly {Robust} {Proximal} {Causal} {Inference} under {Confounded} {Outcome}-{Dependent} {Sampling}},
    url = {http://arxiv.org/abs/2208.01237},
    doi = {10.48550/arXiv.2208.01237},
    urldate = {2025-05-20},
    publisher = {arXiv},
    author = {Li, Kendrick Qijun and Shi, Xu and Miao, Wang and Tchetgen Tchetgen, Eric},
    month = aug,
    year = {2022},
    note = {arXiv:2208.01237 [stat]},
}

@misc{wu_doubly_2023,
    title = {Doubly {Robust} {Proximal} {Causal} {Learning} for {Continuous} {Treatments}},
    copyright = {arXiv.org perpetual, non-exclusive license},
    url = {https://arxiv.org/abs/2309.12819},
    doi = {10.48550/ARXIV.2309.12819},
    urldate = {2025-05-20},
    publisher = {arXiv},
    author = {Wu, Yong and Fu, Yanwei and Wang, Shouyan and Sun, Xinwei},
    year = {2023},
    note = {Version Number: 3},
}

@misc{chen_proximal_2024,
    title = {Proximal {Causal} {Inference} {With} {Text} {Data}},
    url = {http://arxiv.org/abs/2401.06687},
    doi = {10.48550/arXiv.2401.06687},
    urldate = {2025-04-07},
    publisher = {arXiv},
    author = {Chen, Jacob M. and Bhattacharya, Rohit and Keith, Katherine A.},
    month = oct,
    year = {2024},
    note = {arXiv:2401.06687 [cs]},
}

@article{liu_introduction_2013,
    title = {An {Introduction} to {Sensitivity} {Analysis} for {Unobserved} {Confounding} in {Nonexperimental} {Prevention} {Research}},
    volume = {14},
    copyright = {http://www.springer.com/tdm},
    issn = {1389-4986, 1573-6695},
    url = {http://link.springer.com/10.1007/s11121-012-0339-5},
    doi = {10.1007/s11121-012-0339-5},
    language = {en},
    number = {6},
    urldate = {2025-01-29},
    journal = {Prevention Science},
    author = {Liu, Weiwei and Kuramoto, S. Janet and Stuart, Elizabeth A.},
    month = dec,
    year = {2013},
    pages = {570--580},
}

@article{hong_bayesian_2017,
    title = {Bayesian {Approach} for {Addressing} {Differential} {Covariate} {Measurement} {Error} in {Propensity} {Score} {Methods}},
    volume = {82},
    issn = {0033-3123, 1860-0980},
    url = {http://link.springer.com/10.1007/s11336-016-9533-x},
    doi = {10.1007/s11336-016-9533-x},
    language = {en},
    number = {4},
    urldate = {2024-07-16},
    journal = {Psychometrika},
    author = {Hong, Hwanhee and Rudolph, Kara E. and Stuart, Elizabeth A.},
    month = dec,
    year = {2017},
    pages = {1078--1096},
}

@article{zivich_introducing_2023,
    title = {{INTRODUCING} {PROXIMAL} {CAUSAL} {INFERENCE} {FOR} {EPIDEMIOLOGISTS}},
    volume = {192},
    copyright = {https://academic.oup.com/pages/standard-publication-reuse-rights},
    issn = {0002-9262, 1476-6256},
    url = {https://academic.oup.com/aje/article/192/7/1224/7098281},
    doi = {10.1093/aje/kwad077},
    language = {en},
    number = {7},
    urldate = {2025-05-19},
    journal = {American Journal of Epidemiology},
    author = {Zivich, Paul N and Cole, Stephen R and Edwards, Jessie K and Mulholland, Grace E and Shook-Sa, Bonnie E and Tchetgen Tchetgen, Eric J},
    month = jul,
    year = {2023},
    pages = {1224--1227},
}

@article{shi_multiply_2020,
    title = {Multiply {Robust} {Causal} {Inference} with {Double}-{Negative} {Control} {Adjustment} for {Categorical} {Unmeasured} {Confounding}},
    volume = {82},
    copyright = {https://academic.oup.com/journals/pages/open\_access/funder\_policies/chorus/standard\_publication\_model},
    issn = {1369-7412, 1467-9868},
    url = {https://academic.oup.com/jrsssb/article/82/2/521/7056052},
    doi = {10.1111/rssb.12361},
    language = {en},
    number = {2},
    urldate = {2025-03-24},
    journal = {Journal of the Royal Statistical Society Series B: Statistical Methodology},
    author = {Shi, Xu and Miao, Wang and Nelson, Jennifer C. and Tchetgen Tchetgen, Eric J.},
    month = apr,
    year = {2020},
    pages = {521--540},
}

@article{cui_semiparametric_2024,
    title = {Semiparametric {Proximal} {Causal} {Inference}},
    volume = {119},
    issn = {0162-1459, 1537-274X},
    url = {https://www.tandfonline.com/doi/full/10.1080/01621459.2023.2191817},
    doi = {10.1080/01621459.2023.2191817},
    language = {en},
    number = {546},
    urldate = {2024-09-26},
    journal = {Journal of the American Statistical Association},
    author = {Cui, Yifan and Pu, Hongming and Shi, Xu and Miao, Wang and Tchetgen Tchetgen, Eric},
    month = apr,
    year = {2024},
    pages = {1348--1359},
}

@article{miao_identifying_2023,
    title = {Identifying {Effects} of {Multiple} {Treatments} in the {Presence} of {Unmeasured} {Confounding}},
    volume = {118},
    issn = {0162-1459, 1537-274X},
    url = {https://www.tandfonline.com/doi/full/10.1080/01621459.2021.2023551},
    doi = {10.1080/01621459.2021.2023551},
    language = {en},
    number = {543},
    urldate = {2025-04-22},
    journal = {Journal of the American Statistical Association},
    author = {Miao, Wang and Hu, Wenjie and Ogburn, Elizabeth L. and Zhou, Xiao-Hua},
    month = jul,
    year = {2023},
    pages = {1953--1967},
}

@misc{liu_regression-based_2024,
    title = {Regression-{Based} {Proximal} {Causal} {Inference}},
    copyright = {arXiv.org perpetual, non-exclusive license},
    url = {https://arxiv.org/abs/2402.00335},
    doi = {10.48550/ARXIV.2402.00335},
    urldate = {2024-09-26},
    publisher = {arXiv},
    author = {Liu, Jiewen and Park, Chan and Li, Kendrick and Tchetgen Tchetgen, Eric J.},
    year = {2024},
    note = {Version Number: 3},
}

@misc{ghassami_minimax_2021,
    title = {Minimax {Kernel} {Machine} {Learning} for a {Class} of {Doubly} {Robust} {Functionals} with {Application} to {Proximal} {Causal} {Inference}},
    copyright = {Creative Commons Attribution 4.0 International},
    url = {https://arxiv.org/abs/2104.02929},
    doi = {10.48550/ARXIV.2104.02929},
    urldate = {2025-04-16},
    publisher = {arXiv},
    author = {Ghassami, AmirEmad and Ying, Andrew and Shpitser, Ilya and Tchetgen Tchetgen, Eric},
    year = {2021},
    note = {Version Number: 3},
}

@article{newey_instrumental_2003,
    title = {Instrumental {Variable} {Estimation} of {Nonparametric} {Models}},
    volume = {71},
    copyright = {http://doi.wiley.com/10.1002/tdm\_license\_1.1},
    issn = {0012-9682, 1468-0262},
    url = {http://doi.wiley.com/10.1111/1468-0262.00459},
    doi = {10.1111/1468-0262.00459},
    language = {en},
    number = {5},
    urldate = {2025-08-20},
    journal = {Econometrica},
    author = {Newey, Whitney K. and Powell, James L.},
    month = sep,
    year = {2003},
    pages = {1565--1578},
}

@misc{rakshit_adaptive_2025,
    title = {Adaptive {Proximal} {Causal} {Inference} with {Some} {Invalid} {Proxies}},
    url = {http://arxiv.org/abs/2507.19623},
    doi = {10.48550/arXiv.2507.19623},
    urldate = {2025-08-27},
    publisher = {arXiv},
    author = {Rakshit, Prabrisha and Shi, Xu and Tchetgen Tchetgen, Eric},
    month = jul,
    year = {2025},
    note = {arXiv:2507.19623 [stat]},
}

@misc{yu_fortified_2025,
    title = {Fortified {Proximal} {Causal} {Inference} with {Many} {Invalid} {Proxies}},
    url = {http://arxiv.org/abs/2506.13152},
    doi = {10.48550/arXiv.2506.13152},
    urldate = {2025-08-27},
    publisher = {arXiv},
    author = {Yu, Myeonghun and Shi, Xu and Tchetgen Tchetgen, Eric J.},
    month = jun,
    year = {2025},
    note = {arXiv:2506.13152 [stat]},
}

@article{innes_measurement_2021,
    title = {The {Measurement} {Error} {Elephant} in the {Room}: {Challenges} and {Solutions} to {Measurement} {Error} in {Epidemiology}},
    volume = {43},
    copyright = {https://academic.oup.com/journals/pages/open\_access/funder\_policies/chorus/standard\_publication\_model},
    issn = {1478-6729},
    shorttitle = {The {Measurement} {Error} {Elephant} in the {Room}},
    url = {https://academic.oup.com/epirev/article/43/1/94/6401584},
    doi = {10.1093/epirev/mxab011},
    language = {en},
    number = {1},
    urldate = {2025-03-21},
    journal = {Epidemiologic Reviews},
    author = {Innes, Gabriel K and Bhondoekhan, Fiona and Lau, Bryan and Gross, Alden L and Ng, Derek K and Abraham, Alison G},
    month = dec,
    year = {2021},
    pages = {94--105},
}

@article{zafari_state_2024,
    title = {The {State} of {Use} and {Utility} of {Negative} {Controls} in {Pharmacoepidemiologic} {Studies}},
    volume = {193},
    copyright = {https://academic.oup.com/journals/pages/open\_access/funder\_policies/chorus/standard\_publication\_model},
    issn = {0002-9262, 1476-6256},
    url = {https://academic.oup.com/aje/article/193/3/426/7320297},
    doi = {10.1093/aje/kwad201},
    language = {en},
    number = {3},
    urldate = {2025-09-09},
    journal = {American Journal of Epidemiology},
    author = {Zafari, Zafar and Park, Jeong-eun and Shah, Chintal H and dosReis, Susan and Gorman, Emily F and Hua, Wei and Ma, Yong and Tian, Fang},
    month = mar,
    year = {2024},
    pages = {426--453},
}

@article{shi_selective_2020,
    title = {A {Selective} {Review} of {Negative} {Control} {Methods} in {Epidemiology}},
    volume = {7},
    issn = {2196-2995},
    url = {https://link.springer.com/10.1007/s40471-020-00243-4},
    doi = {10.1007/s40471-020-00243-4},
    language = {en},
    number = {4},
    urldate = {2025-03-01},
    journal = {Current Epidemiology Reports},
    author = {Shi, Xu and Miao, Wang and Tchetgen Tchetgen, Eric},
    month = dec,
    year = {2020},
    pages = {190--202},
}

@article{baiocchi_instrumental_2014,
    title = {Instrumental variable methods for causal inference: {Instrumental} variable methods for causal inference},
    volume = {33},
    issn = {02776715},
    shorttitle = {Instrumental variable methods for causal inference},
    url = {https://onlinelibrary.wiley.com/doi/10.1002/sim.6128},
    doi = {10.1002/sim.6128},
    language = {en},
    number = {13},
    urldate = {2023-08-06},
    journal = {Statistics in Medicine},
    author = {Baiocchi, Michael and Cheng, Jing and Small, Dylan S.},
    month = jun,
    year = {2014},
    pages = {2297--2340},
}

@article{wang_matching_2003,
    title = {Matching conditional and marginal shapes in binary random intercept models using a bridge distribution function},
    volume = {90},
    issn = {0006-3444, 1464-3510},
    url = {https://academic.oup.com/biomet/article-lookup/doi/10.1093/biomet/90.4.765},
    doi = {10.1093/biomet/90.4.765},
    language = {en},
    number = {4},
    urldate = {2025-09-30},
    journal = {Biometrika},
    author = {Wang, Z.},
    month = dec,
    year = {2003},
    pages = {765--775},
}

@article{miao_confounding_2024,
    title = {A confounding bridge approach for double negative control inference on causal effects},
    volume = {8},
    issn = {2475-4269, 2475-4277},
    url = {https://www.tandfonline.com/doi/full/10.1080/24754269.2024.2390748},
    doi = {10.1080/24754269.2024.2390748},
    language = {en},
    number = {4},
    urldate = {2025-01-07},
    journal = {Statistical Theory and Related Fields},
    author = {Miao, Wang and Shi, Xu and Li, Yilin and Tchetgen Tchetgen, Eric J.},
    month = oct,
    year = {2024},
    pages = {262--273},
}

@article{bound_problems_1995,
    title = {Problems with {Instrumental} {Variables} {Estimation} when the {Correlation} between the {Instruments} and the {Endogenous} {Explanatory} {Variable} is {Weak}},
    volume = {90},
    issn = {0162-1459, 1537-274X},
    url = {http://www.tandfonline.com/doi/abs/10.1080/01621459.1995.10476536},
    doi = {10.1080/01621459.1995.10476536},
    language = {en},
    number = {430},
    urldate = {2025-10-02},
    journal = {Journal of the American Statistical Association},
    author = {Bound, John and Jaeger, David A. and Baker, Regina M.},
    month = jun,
    year = {1995},
    pages = {443--450},
}

@article{angrist_identification_1996,
    title = {Identification of {Causal} {Effects} {Using} {Instrumental} {Variables}},
    volume = {91},
    issn = {0162-1459, 1537-274X},
    url = {http://www.tandfonline.com/doi/abs/10.1080/01621459.1996.10476902},
    doi = {10.1080/01621459.1996.10476902},
    language = {en},
    number = {434},
    urldate = {2025-10-02},
    journal = {Journal of the American Statistical Association},
    author = {Angrist, Joshua D. and Imbens, Guido W. and Rubin, Donald B.},
    month = jun,
    year = {1996},
    pages = {444--455},
}

@article{yang_advances_2024,
    title = {Advances in methodologies of negative controls: a scoping review},
    volume = {166},
    issn = {08954356},
    shorttitle = {Advances in methodologies of negative controls},
    url = {https://linkinghub.elsevier.com/retrieve/pii/S0895435623003189},
    doi = {10.1016/j.jclinepi.2023.111228},
    language = {en},
    urldate = {2024-08-06},
    journal = {Journal of Clinical Epidemiology},
    author = {Yang, Qingqing and Yang, Zhirong and Cai, Xianming and Zhao, Houyu and Jia, Jinzhu and Sun, Feng},
    month = feb,
    year = {2024},
    pages = {111228},
}

@article{rosenbaum_assessing_1983,
    title = {Assessing {Sensitivity} to an {Unobserved} {Binary} {Covariate} in an {Observational} {Study} with {Binary} {Outcome}},
    volume = {45},
    copyright = {https://academic.oup.com/journals/pages/open\_access/funder\_policies/chorus/standard\_publication\_model},
    issn = {1369-7412, 1467-9868},
    url = {https://academic.oup.com/jrsssb/article/45/2/212/7027731},
    doi = {10.1111/j.2517-6161.1983.tb01242.x},
    language = {en},
    number = {2},
    urldate = {2025-10-02},
    journal = {Journal of the Royal Statistical Society Series B: Statistical Methodology},
    author = {Rosenbaum, P. R. and Rubin, D. B.},
    month = jan,
    year = {1983},
    pages = {212--218},
}

@article{sturmer_adjusting_2005,
    title = {Adjusting {Effect} {Estimates} for {Unmeasured} {Confounding} with {Validation} {Data} using {Propensity} {Score} {Calibration}},
    volume = {162},
    issn = {1476-6256, 0002-9262},
    url = {http://academic.oup.com/aje/article/162/3/279/171198/Adjusting-Effect-Estimates-for-Unmeasured},
    doi = {10.1093/aje/kwi192},
    language = {en},
    number = {3},
    urldate = {2025-01-06},
    journal = {American Journal of Epidemiology},
    author = {Stürmer, Til and Schneeweiss, Sebastian and Avorn, Jerry and Glynn, Robert J.},
    month = aug,
    year = {2005},
    pages = {279--289},
}

@article{nguyen_propensity_2020,
    title = {Propensity {Score} {Analysis} {With} {Latent} {Covariates}: {Measurement} {Error} {Bias} {Correction} {Using} the {Covariate}’s {Posterior} {Mean}, aka the \textit{{Inclusive}} {Factor} {Score}},
    volume = {45},
    issn = {1076-9986, 1935-1054},
    shorttitle = {Propensity {Score} {Analysis} {With} {Latent} {Covariates}},
    url = {http://journals.sagepub.com/doi/10.3102/1076998620911920},
    doi = {10.3102/1076998620911920},
    language = {en},
    number = {5},
    urldate = {2024-09-12},
    journal = {Journal of Educational and Behavioral Statistics},
    author = {Nguyen, Trang Quynh and Stuart, Elizabeth A.},
    month = oct,
    year = {2020},
    pages = {598--636},
}

@article{rubin_estimating_1974,
    title = {Estimating causal effects of treatments in randomized and nonrandomized studies.},
    volume = {66},
    issn = {1939-2176, 0022-0663},
    url = {https://doi.apa.org/doi/10.1037/h0037350},
    doi = {10.1037/h0037350},
    language = {en},
    number = {5},
    urldate = {2025-01-02},
    journal = {Journal of Educational Psychology},
    author = {Rubin, Donald B.},
    month = oct,
    year = {1974},
    pages = {688--701},
}

@article{rubin_randomization_1980,
    title = {Randomization {Analysis} of {Experimental} {Data}: {The} {Fisher} {Randomization} {Test} {Comment}},
    volume = {75},
    issn = {01621459},
    shorttitle = {Randomization {Analysis} of {Experimental} {Data}},
    url = {https://www.jstor.org/stable/2287653?origin=crossref},
    doi = {10.2307/2287653},
    number = {371},
    urldate = {2025-01-02},
    journal = {Journal of the American Statistical Association},
    author = {Rubin, Donald B.},
    month = sep,
    year = {1980},
    pages = {591},
}

@article{bang_doubly_2005,
    title = {Doubly {Robust} {Estimation} in {Missing} {Data} and {Causal} {Inference} {Models}},
    volume = {61},
    copyright = {http://onlinelibrary.wiley.com/termsAndConditions\#vor},
    issn = {0006-341X, 1541-0420},
    url = {https://academic.oup.com/biometrics/article/61/4/962-973/7296220},
    doi = {10.1111/j.1541-0420.2005.00377.x},
    language = {en},
    number = {4},
    urldate = {2025-06-17},
    journal = {Biometrics},
    author = {Bang, Heejung and Robins, James M.},
    month = dec,
    year = {2005},
    pages = {962--973},
}

@book{carroll_measurement_2006,
    edition = {0},
    title = {Measurement {Error} in {Nonlinear} {Models}},
    isbn = {978-1-4200-1013-8},
    url = {https://www.taylorfrancis.com/books/9781420010138},
    language = {en},
    urldate = {2025-06-12},
    publisher = {Chapman and Hall/CRC},
    author = {Carroll, Raymond J. and Ruppert, David and Stefanski, Leonard A. and Crainiceanu, Ciprian M.},
    month = jun,
    year = {2006},
    doi = {10.1201/9781420010138},
}

@misc{mastouri_proximal_2023,
    title = {Proximal {Causal} {Learning} with {Kernels}: {Two}-{Stage} {Estimation} and {Moment} {Restriction}},
    shorttitle = {Proximal {Causal} {Learning} with {Kernels}},
    url = {http://arxiv.org/abs/2105.04544},
    doi = {10.48550/arXiv.2105.04544},
    urldate = {2025-04-07},
    publisher = {arXiv},
    author = {Mastouri, Afsaneh and Zhu, Yuchen and Gultchin, Limor and Korba, Anna and Silva, Ricardo and Kusner, Matt J. and Gretton, Arthur and Muandet, Krikamol},
    month = mar,
    year = {2023},
    note = {arXiv:2105.04544 [cs]},
}

@article{hernan_target_2022,
    title = {Target {Trial} {Emulation}: {A} {Framework} for {Causal} {Inference} {From} {Observational} {Data}},
    volume = {328},
    issn = {0098-7484},
    shorttitle = {Target {Trial} {Emulation}},
    url = {https://jamanetwork.com/journals/jama/fullarticle/2799678},
    doi = {10.1001/jama.2022.21383},
    language = {en},
    number = {24},
    urldate = {2024-08-05},
    journal = {JAMA},
    author = {Hernán, Miguel A. and Wang, Wei and Leaf, David E.},
    month = dec,
    year = {2022},
    pages = {2446},
}

@article{mattingly_effectiveness_2023,
    title = {Effectiveness of vortioxetine for major depressive disorder in real-world clinical practice: {US} cohort results from the global {RELIEVE} study},
    volume = {13},
    issn = {1664-0640},
    shorttitle = {Effectiveness of vortioxetine for major depressive disorder in real-world clinical practice},
    url = {https://www.frontiersin.org/articles/10.3389/fpsyt.2022.977560/full},
    doi = {10.3389/fpsyt.2022.977560},
    urldate = {2025-09-05},
    journal = {Frontiers in Psychiatry},
    author = {Mattingly, Gregory and Brunner, Elizabeth and Chrones, Lambros and Lawrence, Debra F. and Simonsen, Kenneth and Ren, Hongye},
    month = jan,
    year = {2023},
    pages = {977560},
}

@article{rush_acute_2006,
    title = {Acute and {Longer}-{Term} {Outcomes} in {Depressed} {Outpatients} {Requiring} {One} or {Several} {Treatment} {Steps}: {A} {STAR}*{D} {Report}},
    volume = {163},
    issn = {0002-953X, 1535-7228},
    shorttitle = {Acute and {Longer}-{Term} {Outcomes} in {Depressed} {Outpatients} {Requiring} {One} or {Several} {Treatment} {Steps}},
    url = {https://psychiatryonline.org/doi/10.1176/ajp.2006.163.11.1905},
    doi = {10.1176/ajp.2006.163.11.1905},
    language = {en},
    number = {11},
    urldate = {2025-01-14},
    journal = {American Journal of Psychiatry},
    author = {Rush, A. John and Trivedi, Madhukar H. and Wisniewski, Stephen R. and Nierenberg, Andrew A. and Stewart, Jonathan W. and Warden, Diane and Niederehe, George and Thase, Michael E. and Lavori, Philip W. and Lebowitz, Barry D. and McGrath, Patrick J. and Rosenbaum, Jerrold F. and Sackeim, Harold A. and Kupfer, David J. and Luther, James and Fava, Maurizio},
    month = nov,
    year = {2006},
    pages = {1905--1917},
}

@article{lin_out--system_2018,
    title = {Out-of-system {Care} and {Recording} of {Patient} {Characteristics} {Critical} for {Comparative} {Effectiveness} {Research}:},
    volume = {29},
    issn = {1044-3983},
    shorttitle = {Out-of-system {Care} and {Recording} of {Patient} {Characteristics} {Critical} for {Comparative} {Effectiveness} {Research}},
    url = {http://journals.lww.com/00001648-201805000-00006},
    doi = {10.1097/EDE.0000000000000794},
    language = {en},
    number = {3},
    urldate = {2025-08-28},
    journal = {Epidemiology},
    author = {Lin, Kueiyu Joshua and Glynn, Robert J. and Singer, Daniel E. and Murphy, Shawn N. and Lii, Joyce and Schneeweiss, Sebastian},
    month = may,
    year = {2018},
    pages = {356--363},
}

@article{gronemann_incidence_2018,
    title = {Incidence of, {Risk} {Factors} for, and {Changes} {Over} {Time} in {Treatment}-{Resistant} {Depression} in {Denmark}: {A} {Register}-{Based} {Cohort} {Study}},
    volume = {79},
    issn = {1555-2101},
    shorttitle = {Incidence of, {Risk} {Factors} for, and {Changes} {Over} {Time} in {Treatment}-{Resistant} {Depression} in {Denmark}},
    url = {https://www.psychiatrist.com/jcp/trd-incidence-risk-factors-and-change-over-time},
    doi = {10.4088/JCP.17m11845},
    number = {4},
    urldate = {2025-09-05},
    journal = {The Journal of Clinical Psychiatry},
    author = {Gronemann, Frederikke Hordam and Jorgensen, Martin B. and Nordentoft, Merete and Andersen, Per K. and Osler, Merete},
    month = may,
    year = {2018},
}

@misc{xie_automating_2024,
    title = {Automating the {Selection} of {Proxy} {Variables} of {Unmeasured} {Confounders}},
    url = {http://arxiv.org/abs/2405.16130},
    urldate = {2024-07-16},
    publisher = {arXiv},
    author = {Xie, Feng and Chen, Zhengming and Luo, Shanshan and Miao, Wang and Cai, Ruichu and Geng, Zhi},
    month = may,
    year = {2024},
    note = {arXiv:2405.16130 [cs, stat]},
}

@misc{kummerfeld_data-driven_2022,
    title = {Data-driven {Automated} {Negative} {Control} {Estimation} ({DANCE}): {Search} for, {Validation} of, and {Causal} {Inference} with {Negative} {Controls}},
    copyright = {Creative Commons Attribution 4.0 International},
    shorttitle = {Data-driven {Automated} {Negative} {Control} {Estimation} ({DANCE})},
    url = {https://arxiv.org/abs/2210.00528},
    doi = {10.48550/ARXIV.2210.00528},
    urldate = {2025-01-10},
    publisher = {arXiv},
    author = {Kummerfeld, Erich and Lim, Jaewon and Shi, Xu},
    year = {2022},
    note = {Version Number: 1},
}

@misc{ying_proximal_2021,
    title = {Proximal {Causal} {Inference} for {Complex} {Longitudinal} {Studies}},
    copyright = {arXiv.org perpetual, non-exclusive license},
    url = {https://arxiv.org/abs/2109.07030},
    doi = {10.48550/ARXIV.2109.07030},
    urldate = {2025-10-08},
    publisher = {arXiv},
    author = {Ying, Andrew and Miao, Wang and Shi, Xu and Tchetgen Tchetgen, Eric J.},
    year = {2021},
    note = {Version Number: 5},
}

@incollection{kline_chapter_2016,
    address = {New York London},
    edition = {Fourth edition},
    series = {Methodology in the social sciences},
    title = {Chapter 9: {Specification} and {Identification} of {Confirmatory} {Factor} {Analysis} {Models}},
    isbn = {978-1-4625-2335-1 978-1-4625-2336-8},
    language = {eng},
    booktitle = {Principles and practice of structural equation modeling},
    publisher = {The Guilford Press},
    author = {Kline, Rex B.},
    year = {2016},
}

@misc{cobzaru_bias_2022,
    title = {Bias {Formulas} for {Violations} of {Proximal} {Identification} {Assumptions}},
    url = {http://arxiv.org/abs/2208.00105},
    doi = {10.48550/arXiv.2208.00105},
    urldate = {2025-05-06},
    publisher = {arXiv},
    author = {Cobzaru, Raluca and Welsch, Roy and Finkelstein, Stan and Ng, Kenney and Shahn, Zach},
    month = aug,
    year = {2022},
    note = {arXiv:2208.00105 [math]},
}

@article{rudolph_using_2018,
    title = {Using {Sensitivity} {Analyses} for {Unobserved} {Confounding} to {Address} {Covariate} {Measurement} {Error} in {Propensity} {Score} {Methods}},
    volume = {187},
    issn = {0002-9262, 1476-6256},
    url = {https://academic.oup.com/aje/article/187/3/604/3964398},
    doi = {10.1093/aje/kwx248},
    language = {en},
    number = {3},
    urldate = {2024-07-16},
    journal = {American Journal of Epidemiology},
    author = {Rudolph, Kara E and Stuart, Elizabeth A},
    month = mar,
    year = {2018},
    pages = {604--613},
}

@article{vanderweele_bias_2011,
    title = {Bias {Formulas} for {Sensitivity} {Analysis} of {Unmeasured} {Confounding} for {General} {Outcomes}, {Treatments}, and {Confounders}},
    volume = {22},
    issn = {1044-3983},
    url = {https://journals.lww.com/00001648-201101000-00008},
    doi = {10.1097/EDE.0b013e3181f74493},
    language = {en},
    number = {1},
    urldate = {2024-09-05},
    journal = {Epidemiology},
    author = {VanderWeele, Tyler J. and Arah, Onyebuchi A.},
    month = jan,
    year = {2011},
    pages = {42--52},
}

@misc{ghassami_partial_2023,
    title = {Partial {Identification} of {Causal} {Effects} {Using} {Proxy} {Variables}},
    copyright = {Creative Commons Attribution 4.0 International},
    url = {https://arxiv.org/abs/2304.04374},
    doi = {10.48550/ARXIV.2304.04374},
    urldate = {2024-08-06},
    publisher = {arXiv},
    author = {Ghassami, AmirEmad and Shpitser, Ilya and Tchetgen Tchetgen, Eric},
    year = {2023},
    note = {Version Number: 3},
}

@misc{park_proximal_2025,
    title = {Proximal {Causal} {Inference} for {Conditional} {Separable} {Effects}},
    url = {http://arxiv.org/abs/2402.11020},
    doi = {10.48550/arXiv.2402.11020},
    urldate = {2025-10-21},
    publisher = {arXiv},
    author = {Park, Chan and Stensrud, Mats and Tchetgen, Eric Tchetgen},
    month = apr,
    year = {2025},
    note = {arXiv:2402.11020 [stat]},
}

@misc{liu_proximal_2023,
    title = {Proximal {Causal} {Inference} for {Synthetic} {Control} with {Surrogates}},
    url = {http://arxiv.org/abs/2308.09527},
    doi = {10.48550/arXiv.2308.09527},
    urldate = {2025-10-21},
    publisher = {arXiv},
    author = {Liu, Jizhou and Tchetgen, Eric J. Tchetgen and Varjão, Carlos},
    month = aug,
    year = {2023},
    note = {arXiv:2308.09527 [stat]},
}

@article{zhang_proximal_2023,
    title = {Proximal causal inference without uniqueness assumptions},
    volume = {198},
    issn = {01677152},
    url = {https://linkinghub.elsevier.com/retrieve/pii/S0167715223000603},
    doi = {10.1016/j.spl.2023.109836},
    language = {en},
    urldate = {2025-05-20},
    journal = {Statistics \& Probability Letters},
    author = {Zhang, Jeffrey and Li, Wei and Miao, Wang and Tchetgen Tchetgen, Eric},
    month = jul,
    year = {2023},
    pages = {109836},
}

@Manual{pci2s_package,
    title = {pci2s: Two-Stage-Least-Square Method for Proximal Causal Inference},
    author = {Kendrick Li},
    year = {2025},
    note = {R package version 0.1.0},
    url = {https://github.com/KenLi93/pci2s},
  }

@article{canay_testability_2013,
    title = {On the {Testability} of {Identification} in {Some} {Nonparametric} {Models} {With} {Endogeneity}},
    volume = {81},
    copyright = {http://doi.wiley.com/10.1002/tdm\_license\_1.1},
    issn = {0012-9682},
    url = {http://doi.wiley.com/10.3982/ECTA10851},
    doi = {10.3982/ECTA10851},
    language = {en},
    number = {6},
    urldate = {2025-10-08},
    journal = {Econometrica},
    author = {Canay, Ivan A. and Santos, Andre and Shaikh, Azeem},
    year = {2013},
    pages = {2535--2559},
}

@misc{huang_relative_2025,
    title = {Relative {Bias} {Under} {Imperfect} {Identification} in {Observational} {Causal} {Inference}},
    url = {http://arxiv.org/abs/2507.23743},
    doi = {10.48550/arXiv.2507.23743},
    urldate = {2025-10-21},
    publisher = {arXiv},
    author = {Huang, Melody and McCartan, Cory},
    month = jul,
    year = {2025},
    note = {arXiv:2507.23743 [stat]},
}

@article{cornfield_smoking_1959,
    title = {Smoking and lung cancer: recent evidence and a discussion of some questions},
    volume = {22},
    issn = {0027-8874},
    shorttitle = {Smoking and lung cancer},
    language = {eng},
    number = {1},
    journal = {Journal of the National Cancer Institute},
    author = {Cornfield, J. and Haenszel, W. and Hammond, E. C. and Lilienfeld, A. M. and Shimkin, M. B. and Wynder, E. L.},
    month = jan,
    year = {1959},
    pmid = {13621204},
    pages = {173--203},
}

@misc{zhou_causal_2024,
    title = {Causal {Inference} for a {Hidden} {Treatment}},
    copyright = {arXiv.org perpetual, non-exclusive license},
    url = {https://arxiv.org/abs/2405.09080},
    doi = {10.48550/ARXIV.2405.09080},
    urldate = {2025-10-31},
    publisher = {arXiv},
    author = {Zhou, Ying and Tchetgen, Eric Tchetgen},
    year = {2024},
    note = {Version Number: 3},
}

@article{vanderweele_results_2012,
    title = {Results on {Differential} and {Dependent} {Measurement} {Error} of the {Exposure} and the {Outcome} {Using} {Signed} {Directed} {Acyclic} {Graphs}},
    volume = {175},
    issn = {0002-9262, 1476-6256},
    url = {https://academic.oup.com/aje/article-lookup/doi/10.1093/aje/kwr458},
    doi = {10.1093/aje/kwr458},
    language = {en},
    number = {12},
    urldate = {2024-08-30},
    journal = {American Journal of Epidemiology},
    author = {VanderWeele, T. J. and Hernan, M. A.},
    month = jun,
    year = {2012},
    pages = {1303--1310},
}

@book{kress_linear_2014,
    address = {New York, NY},
    edition = {3rd ed. 2014},
    series = {Applied {Mathematical} {Sciences}},
    title = {Linear {Integral} {Equations}},
    isbn = {978-1-4614-9592-5 978-1-4614-9593-2},
    language = {eng},
    number = {82},
    publisher = {Springer},
    author = {Kress, Rainer},
    year = {2014},
    doi = {10.1007/978-1-4614-9593-2},
}

@article{stefanski_calculus_2002,
    title = {The {Calculus} of {M}-{Estimation}},
    volume = {56},
    issn = {0003-1305, 1537-2731},
    url = {http://www.tandfonline.com/doi/abs/10.1198/000313002753631330},
    doi = {10.1198/000313002753631330},
    language = {en},
    number = {1},
    urldate = {2025-11-20},
    journal = {The American Statistician},
    author = {Stefanski, Leonard A and Boos, Dennis D},
    month = feb,
    year = {2002},
    pages = {29--38},
}

@misc{park_single_2024,
    title = {Single {Proxy} {Control}},
    url = {http://arxiv.org/abs/2302.06054},
    language = {en},
    urldate = {2024-08-07},
    publisher = {arXiv},
    author = {Park, Chan and Richardson, David and Tchetgen, Eric Tchetgen},
    month = mar,
    year = {2024},
    note = {arXiv:2302.06054 [stat]},
}

@article{lockwood_matching_2016,
    title = {Matching and {Weighting} {With} {Functions} of {Error}-{Prone} {Covariates} for {Causal} {Inference}},
    volume = {111},
    issn = {0162-1459, 1537-274X},
    url = {https://www.tandfonline.com/doi/full/10.1080/01621459.2015.1122601},
    doi = {10.1080/01621459.2015.1122601},
    language = {en},
    number = {516},
    urldate = {2025-01-09},
    journal = {Journal of the American Statistical Association},
    author = {Lockwood, J. R. and McCaffrey, Daniel F.},
    month = oct,
    year = {2016},
    pages = {1831--1839},
}

@incollection{harris_introduction_1999,
    series = {Themes in {Modern} {Econometrics}},
    title = {Introduction to the {Generalized} {Method} of {Moments} {Estimation}},
    booktitle = {Generalized {Method} of {Moments} {Estimation}},
    publisher = {Cambridge University Press},
    author = {Harris, David and Mátyás, László},
    editor = {Matyas, LaszloEditor},
    year = {1999},
    pages = {3--30},
}

@misc{kallus_causal_2022,
    title = {Causal {Inference} {Under} {Unmeasured} {Confounding} {With} {Negative} {Controls}: {A} {Minimax} {Learning} {Approach}},
    shorttitle = {Causal {Inference} {Under} {Unmeasured} {Confounding} {With} {Negative} {Controls}},
    url = {http://arxiv.org/abs/2103.14029},
    doi = {10.48550/arXiv.2103.14029},
    urldate = {2025-05-20},
    publisher = {arXiv},
    author = {Kallus, Nathan and Mao, Xiaojie and Uehara, Masatoshi},
    month = oct,
    year = {2022},
    note = {arXiv:2103.14029 [stat]},
}

@misc{sverdrup_proximal_2023,
    title = {Proximal {Causal} {Learning} of {Conditional} {Average} {Treatment} {Effects}},
    copyright = {arXiv.org perpetual, non-exclusive license},
    url = {https://arxiv.org/abs/2301.10913},
    doi = {10.48550/ARXIV.2301.10913},
    urldate = {2025-11-24},
    publisher = {arXiv},
    author = {Sverdrup, Erik and Cui, Yifan},
    year = {2023},
    note = {Version Number: 2},
}

@incollection{newey_chapter_1994,
    title = {Chapter 36 {Large} sample estimation and hypothesis testing},
    volume = {4},
    isbn = {978-0-444-88766-5},
    url = {https://linkinghub.elsevier.com/retrieve/pii/S1573441205800054},
    language = {en},
    urldate = {2025-11-25},
    booktitle = {Handbook of {Econometrics}},
    publisher = {Elsevier},
    author = {Newey, Whitney K. and McFadden, Daniel},
    year = {1994},
    doi = {10.1016/S1573-4412(05)80005-4},
    pages = {2111--2245},
}

@misc{chernozhukov_doubledebiased_2024,
    title = {Double/{Debiased} {Machine} {Learning} for {Treatment} and {Causal} {Parameters}},
    url = {http://arxiv.org/abs/1608.00060},
    doi = {10.48550/arXiv.1608.00060},
    language = {en},
    urldate = {2025-04-15},
    publisher = {arXiv},
    author = {Chernozhukov, Victor and Chetverikov, Denis and Demirer, Mert and Duflo, Esther and Hansen, Christian and Newey, Whitney and Robins, James},
    month = nov,
    year = {2024},
    note = {arXiv:1608.00060 [stat]},
}

@article{andrews_examples_2017,
    title = {Examples of {L} 2 -complete and boundedly-complete distributions},
    volume = {199},
    issn = {03044076},
    url = {https://linkinghub.elsevier.com/retrieve/pii/S0304407617300738},
    doi = {10.1016/j.jeconom.2017.05.011},
    language = {en},
    number = {2},
    urldate = {2026-01-16},
    journal = {Journal of Econometrics},
    author = {Andrews, Donald W.K.},
    month = aug,
    year = {2017},
    pages = {213--220},
}

@article{hu_nonparametric_2018,
    title = {{NONPARAMETRIC} {IDENTIFICATION} {USING} {INSTRUMENTAL} {VARIABLES}: {SUFFICIENT} {CONDITIONS} {FOR} {COMPLETENESS}},
    volume = {34},
    copyright = {https://www.cambridge.org/core/terms},
    issn = {0266-4666, 1469-4360},
    shorttitle = {{NONPARAMETRIC} {IDENTIFICATION} {USING} {INSTRUMENTAL} {VARIABLES}},
    url = {https://www.cambridge.org/core/product/identifier/S0266466617000251/type/journal_article},
    doi = {10.1017/S0266466617000251},
    language = {en},
    number = {3},
    urldate = {2026-01-16},
    journal = {Econometric Theory},
    author = {Hu, Yingyao and Shiu, Ji-Liang},
    month = jun,
    year = {2018},
    pages = {659--693},
}

\end{document}